\documentclass[epj]{svjour}
\usepackage{graphics}
\usepackage{xcolor}
\usepackage[normalem]{ulem}  
\begin{document}
\title{Parametrisations of relativistic energy density functionals with tensor couplings}
\titlerunning{Parametrisations of relativistic energy density functionals with tensor couplings}
\author{Stefan Typel\inst{1,2} \and Diana Alvear Terrero\inst{3,4}
}                     
\authorrunning{S. Typel, D. Alvear Terrero}
%
%
\institute{Technische Universit\"{a}t Darmstadt,
  Fachbereich Physik,
  Institut f\"{u}r Kernphysik, Schlossgartenstra\ss{}e 9,
  64289 Darmstadt, Germany \and
  GSI Helmholtzzentrum f\"{u}r Schwerionenforschung GmbH,
  Theorie,
  Planckstra\ss{}e 1,
  64291 Darmstadt, Germany \and
  Uniwersytet Wroc\l{}awski,
  Instytut Fizyki Teoretycznej,
  pl. M. Borna 9,
  50-204 Wroc\l{}aw, Poland \and
  Instituto de Cibern\'{e}tica Matem\'{a}tica y F\'{i}sica,
  Departamento de F\'{i}sica Te\'{o}rica,
  Calle E esq 15 No. 309,
  Vedado,
  La Habana 10400, Cuba}
\date{Received: date / Revised version: date}
%
\abstract{
  The relativistic density functional with minimal density dependent
  nucleon-meson couplings for nuclei and nuclear matter is extended to
  include tensor couplings of the nucleons to the vector mesons.
  The dependence of the minimal couplings on either vector or scalar densities
  is explored. New parametrisations are obtained by a fit to nuclear observables
  with uncertainties that are determined self-consistently. The corresponding
  nuclear matter parameters at saturation are determined including their
  uncertainties. An improvement in the description of nuclear observables,
  in particular for binding energies and diffraction radii,
  is found when tensor couplings are considered, accompanied by an increase
  of the Dirac effective mass.
  The equations of state for symmetric nuclear matter and
  pure neutron matter are studied for all models. The density dependence
  of the nuclear symmetry energy, the Dirac effective masses and
  scalar densities is explored. Problems at high densities
  for parametrisations using a scalar density dependence of the couplings
  are identified due to the rearrangement contributions in the scalar
  self-energies that lead to vanishing Dirac effective masses.
} 
\maketitle
\section{Introduction}
\label{intro}

Since the application of the relativistic mean-field (RMF) approach
in the framework of quantum hadron dynamics,
various kinds of a relativistic energy density functionals (EDFs) have been
developed in the theoretical description of atomic nuclei and nuclear matter,
see, e.g., \cite{Serot:1984ey,Reinhard:1989zi,Ring:1996qi,Meng:2016}.
In these types of phenomenological approaches
the strong interaction between nucleons is described in an effective way
by the exchange of mesons.
In most cases a minimal coupling of mesons to the nucleons is considered with
a strength that is determined by the corresponding coupling constants.
As a result every nucleon $i$  moves in vector ($V_{i}$)
and scalar ($S_{i}$) mean fields. Although the effective
interaction in an EDF shares some similarities with realistic
one-boson exchange (OBE) potentials due to general features of the strong
interaction, it has a more simplistic structural form and cannot describe
the free nucleon-nucleon scattering quantitatively. The model parameters
are treated as independent quantities without assuming constraints
from theoretical considerations, e.g., relations between different couplings
as in OBE potentials.
The coupling strengths between mesons and nucleons in the EDF
are usually obtained by fitting the model predictions of nuclear observables
to experimental data. Sometimes also nuclear matter parameters, which are
extracted indirectly from properties of nuclei
or experiments with heavy-ion collisions, are used as constraints.
In order to obtain a good quantitative description, a medium dependence of the
effective interaction has to be considered. This is most often realized by
nonlinear self-couplings of the mesons or a density dependence of the
nucleon-meson couplings. Many parametrisations \cite{Dutra:2014qga}
have been developed for different applications.
The effects of other types of interaction vertices, e.g., with derivative
couplings \cite{Typel:2002ck,Typel:2005ba,Gaitanos:2009nt,Gaitanos:2011yb,Gaitanos:2011ej,Chen:2012rn,Gaitanos:2012hg,Chen:2014,Gaitanos:2015xpa,Antic:2015tga}
or tensor couplings, are less intensively explored.

Relativistic EDFs are derived originally in the mean-field
or Hartree approximation starting from a covariant Lagrangian density
with nucleon and meson fields as degrees of freedom. This means that the
many-nucleon wave function is a simple product of single particle states
that are occupied according to Fermi-Dirac statistics but an explicit
antisymmetrisation of the many-nucleon wave function, e.g., in form
of a Slater determinant, is not considered. Only later, exchange or Fock
terms were taken into account in the construction of relativistic EDFs,
usually called relativistic Hartree-Fock (RHF) models, see, e.g.,
\cite{Horowitz:1982kp,Bouyssy:1987sh,Bernardos:1993re,Long:2005ne,Long:2008fm}.
In this work the relativistic Hartree approach is utilized because it is
more transparent and less involved from a computational point of view.
Density dependent meson-nucleon couplings in a full RHF model
with finite-range interactions
would also introduce some complications in the derivation of the
field equations that are difficult to treat completely self-consistently
in actual calculations.

The first parametrisation of a RMF model with $\omega$ and $\rho$ tensor
couplings, VT, found from a fit to nuclear observables,
was introduced in \cite{Rufa:1988zz}. A systematic variation
of the coupling strengths showed no strong sensitivity of most observables to
the tensor contributions and
only a slight improvement with a finite $\omega$ tensor coupling.
However, a decrease of the incompressibility $K$ of nuclear matter and
larger effective nucleon masses were found.
The application of RMF calculations with a $\rho$ tensor coupling
was extended from spherical to deformed nuclei in \cite{Jian-Kang:1991hco}.
The dependence of spin-orbit splittings on the strength of the
$\rho$ tensor coupling was explored in \cite{Ren:1995}.
Adding a $\rho$ tensor coupling to existing RMF parametrisations, the
modification of the neutron skin thickness of nuclei was studied later in
\cite{Jiang:2005vu}. Tensor coupling terms were derived
from a systematic expansion in an effective approach using chiral symmetry
in \cite{Furnstahl:1996wv} and two new parametrisations, G1 and G2, were
obtained by a fit to nuclear observables with much stronger $\rho$ tensor
couplings than $\omega$ tensor couplings. The effects on spin-orbit splittings
and the effective mass were studied in \cite{Furnstahl:1997tk}.
Small effects were found in another parametrisation, NL-VT1, determined in
the framework of RMF models with nonlinear
self-couplings of the mesons, which was applied to the study of super-heavy
nuclei \cite{Bender:1999yt}.
The same interaction was used in \cite{Sulaksono:2005ac}
for a comparison of single-particle spectra in comparison to other EDFs.
The modification of nuclear magnetic moments
by isoscalar tensor couplings 
was already investigated in \cite{Bouyssy:1984bg} and a possible
origin in the framework of RMF models was discussed there. Tensor couplings
of nucleons to $\omega$ and $\rho$ mesons in RHF
calculations were considered first in \cite{Bouyssy:1987sh} and several
studies followed. For instance, 
in \cite{Long:2006pn} the influence of tensor couplings
on pseudospin-orbit splittings was investigated, and in \cite{Long:2007dw} 
a new parametrisation, PKA1, for RHF calculations with density dependent
couplings was introduced. The effects on the evolution of shell gaps
and spin-orbits splittings
were explored in \cite{Wang:2012nu} and in \cite{Jiang:2014mca}, respectively,
in comparison to other relativistic EDFs.
Another series of RHF parametrisations was presented in
\cite{Liliani:2016npj} with an extensive exploration of the effects.
The Zimanyi-Moszkowski version of the RMF Lagrangian was generalized
in \cite{Biro:1996nk}
with a $\omega$ tensor coupling contribution to increase the small spin-orbit
splitting of the original model. In spite of all these investigations no
comprehensive understanding of the role and importance of tensor couplings in
relativistic EDFs has been reached and, hence, they are usually not taken
into account in most models.

A specific feature of tensor couplings is the fact that they
contribute to the
strength of the spin-orbit splitting in nuclei. In standard RMF
calculations of spherical nuclei the
size of the energy splitting between single-nucleon levels
of identical orbital angular
momentum $l$ but different total angular momentum $j=l \pm 1/2$ is
tightly correlated with the scalar potential $S_{i}$ and the Dirac effective
mass $m_{i}^{\ast}=m_{i}-S_{i}$ of a nucleon $i$ with vacuum mass $m_{i}$.
The introduction of tensor
couplings can lift this correlation and allows to increase the effective mass
that is usually considered to be smaller than expected to describe the
level density near the Fermi energy.
In calculations of nuclear matter, however, tensor couplings do not contribute
in the conventional mean-field approximation of spatially uniform systems since
their effect depends on spatial derivatives of densities.

It was realized early on that an effective medium dependence of the
interaction has to be included in relativistic approaches
in order to improve the description
of nuclei and nuclear matter quantitatively. One major class of models
considers a self-coupling or cross-coupling of the mesons leading to an
increase of the number of model parameters.
A second class introduces a density dependence of the
meson-nucleon couplings. It can be arbitrarily varied depending on the
selected form of the function.
In addition, a specific feature of relativistic models
can be exploited in this approach.
There are various types of densities and currents
that can be formed in a Lorentz covariant way from the nuclear wave functions
and that can enter as argument in the coupling functions. The most common
approach is a so-called vector density dependence. In contrast, scalar or other
density dependencies are hardly used in recent EDF
parametrisations even though they
were discussed in some of the first applications of the RMF model
\cite{Fuchs:1995as,Lenske:1995wyj}.
Depending on the choice of a vector or scalar
density dependence, so-called rearrangement terms appear in the vector or
scalar mean fields, respectively, in addition to the regular meson
contributions. They are essential for the thermodynamic consistency of the
theory. Differences between models with different density dependencies
seem to be small in the density range where they are constrained
by nuclear data but problems may arise is some parts of the
phase diagram of nuclear matter under particular conditions
\cite{Typel:2018cap}.

In this work a new set of parametrisations for relativistic EDFs
is introduced with a vector or a scalar density dependence
of the couplings and the effects of tensor couplings are studied.
The model parameters are usually
determined by minimising an objective function that depends on the differences
between calculated and measured nuclear data weighted by (inverse)
uncertainties.
The latter are mostly set heuristically to certain fixed values.
These are assumed to be of reasonable size 
matching the expected quality of the model. These uncertainties
are generally larger than the experimental errors of the considered nuclear
data. Here a modified approach is followed.
It allows to adjust the uncertainties
during the determination of the model parameters in the fitting procedure,
see, e.g., \cite{Dobaczewski:2014jga}.
Thus the uncertainties are obtained in a self-consistent way and give a new
possibility to judge the quality of the EDF.

The approach presented here stays on the mean-field level
and does not consider any effects beyond this approximation.
Beyond mean-field effects are known to have a noticeable impact, e.g.,
on single-particle spectra and excited states of nuclei which are,
however, not examined in this work. There are different ways of
treating beyond mean-field effects, e.g., by studying
collective correlations and their fluctuations related to the restoration
of broken symmetries or by applying (quasi-particle)
random phase approximation (RPA) methods
looking at collective phonon states and particle-vibration couplings.
See, e.g., \cite{Niksic:2006kv,Niksic:2006kx,Niksic:2008cn,Niksic:2011sg,Litvinova:2006ds,Litvinova:2007jb,Litvinova:2007gg,Litvinova:2008qy,Litvinova:2008he,Litvinova:2011zz,Litvinova:2014sza,Afanasjev:2014lga,Robin:2016wuh,Arteaga:2009mb,Fu:2013mda,Yao:2014uta,Yao:2015lha} in the context of relativistic models.
Going beyond the mean-field approximation requires usually a complete refit of
the EDF parameters to be consistent. This is computationally very involved and
beyond the scope of the present work. The main aim is to
establish tensor couplings as a useful ingredient
in relativistic EDF calculations.

The content of this study is organised as follows: In section \ref{sec:rEDF}
the theoretical formalism is introduced assuming a density dependence
of the minimal meson-nucleon couplings supplemented with tensor couplings
of the nucleons to $\omega$ and $\rho$ mesons. The specific applications
of the model to spherical nuclei and cold uniform nuclear matter are
discussed in detail. The procedure to determine the model parameters
is outlined in section \ref{sec:para}. This includes the choice of the
coupling functions, the observables and nuclei as well as the definition
of the objective function. Furthermore a set of EDFs is defined that
take different effects into account. A presentation of the results follows
in section \ref{sec:res} with numerical details of various EDF
parametrisations, the corresponding nuclear matter parameters,
figures with the density dependence of various quantities
and a discussion of the fit quality. Conclusion are given
in section \ref{sec:concl}. The derivation of the
rearrangement contributions to the scalar and vector potentials
of the nucleons is presented in appendix A and
the conversion of model parameters is discussed in
appendix B.

\section{Relativistic energy density functional}
\label{sec:rEDF}

The traditional starting point to derive a relativistic EDF
is a Lagrangian density $\mathcal{L}$
with nucleons and mesons as degrees of freedom. Here we follow mostly
the notation in \cite{Typel:2018cap} and include the necessary terms for the
tensor couplings. Natural units with $\hbar = c = 1$ are used throughout.
Neutrons ($i=n$) and protons ($i=p$) are represented by four-spinors
$\Psi_{i}$. Hyperons
or other baryons are not considered in the present work. Usually, four
types of mesons are considered: isoscalar $\omega$ and $\sigma$ mesons
and isovector $\rho$ and $\delta$ mesons where the first of the two pairs
is a Lorentz vector field and the second is a Lorentz scalar. The corresponding
fields carry the same quantum numbers as the experimentally observed
mesons but cannot necessarily be identified with them directly.
In order to have a clear and simple representation,
their fields are denoted with the same symbol as their name, however
Lorentz vector mesons carry an index due to their four-vector nature.
Quantities in bold face can be vectors in coordinate space or in
isospin space depending on the context.
Besides mesons the electromagnetic interaction is included
using the symbol $A^{\mu}$ for the field.
All relevant equations can be derived from $\mathcal{L}$
by standard procedures depending on the employed approximations.

\subsection{Lagrangian density and field equations}

The relativistic Lagrangian density can be written as a sum
\begin{equation}
\label{eq:L}
 \mathcal{L} =  \mathcal{L}_{\rm nucleon} + \mathcal{L}_{\rm meson}
 + \mathcal{L}_{\gamma}
\end{equation}
of three contributions. The first one
\begin{equation}
 \mathcal{L}_{\rm nucleon} = \sum_{i=p,n} \overline{\Psi}_{i}
 \left( \gamma_{\mu}iD_{i}^{\mu} - \sigma_{\mu\nu}T_{i}^{\mu\nu}- M_{i}^{\ast} \right) \Psi_{i}
\end{equation}
contains the nucleon fields $\Psi_{i}$,
$\overline{\Psi}_{i}= \Psi_{i}^{\dagger}\gamma^{0}$
and their coupling to the meson fields
with standard relativistic matrices $\gamma_{\mu}$ and $\sigma_{\mu\nu}$.
In the covariant derivative
\begin{equation}
  \label{eq:D}
  iD_{i}^{\mu} = i \partial^{\mu} - \tilde{\Gamma}_{\omega} \omega^{\mu}
  - \tilde{\Gamma}_{\rho} \vec{\rho}^{\mu} \cdot \vec{\tau}
  - \Gamma_{\gamma}  A^{\mu} \frac{1+\tau_{3}}{2}
\end{equation}
the Lorentz vector mesons $\omega$ and $\rho$ appear.
The isospin matrices $\tau_{k}$ ($k=1,2,3$) are components of the
vector $\vec{\tau}$ in isospin space in analogy
of the Pauli matrices $\sigma_{k}$.
$\Gamma_{\gamma}$ is the electromagnetic coupling
constant and $\tilde{\Gamma}_{j}$
are the meson-nucleon couplings that are functionals of the nucleon fields
$\Psi_{i}$ and $\overline{\Psi}_{i}$. The effective mass
operator for neutrons ($i=n$) and protons ($i=p$) with vacuum rest mass $m_{i}$
\begin{equation}
  \label{eq:M}
 M_{i}^{\ast} =  m_{i} - \tilde{\Gamma}_{\sigma} \sigma 
 - \tilde{\Gamma}_{\delta} \vec{\delta} \cdot \vec{\tau}
\end{equation}
depends on the Lorentz scalar fields $\sigma$ and $\delta$.
Finally, the contribution with
the tensor coupling
\begin{equation}
  T_{i}^{\mu\nu} = \frac{\Gamma_{T\omega}}{2m_{p}} G^{\mu\nu}
  + \frac{\Gamma_{T\rho}}{2m_{p}} \vec{H}^{\mu\nu} \cdot \vec{\tau}
\end{equation}
contains the field tensors 
$G^{\mu\nu}  =  \partial^{\mu} \omega^{\nu} - \partial^{\nu} \omega^{\mu}$
and $\vec{H}^{\mu\nu}  =  \partial^{\mu} \vec{\rho}^{\nu}
- \partial^{\nu} \vec{\rho}^{\mu}$
of the Lorentz vector mesons. Due to the factor $m_{p}$ in the denominator, the
couplings $\Gamma_{T\omega}$ and $\Gamma_{T\rho}$ are dimensionless quantities.
There is no standard convention in the literature on how to write the tensor
coupling term. Thus the size of the couplings is not always
comparable immediately. The term
\begin{eqnarray}
  \lefteqn{\mathcal{L}_{\rm meson}}
  \\ \nonumber & = & \frac{1}{2} \Big( 
 \partial^{\mu} \sigma \partial_{\mu} \sigma - m_{\sigma}^{2} \: \sigma^{2}
 + \partial^{\mu} \vec{\delta} \cdot \partial_{\mu} \vec{\delta} 
 - m_{\delta}^{2} \: \vec{\delta} \cdot \vec{\delta}
 \\ \nonumber & & 
 - \frac{1}{2} G^{\mu\nu} G_{\mu\nu} + m_{\omega}^{2} \: \omega^{\mu}\omega_{\mu}
 - \frac{1}{2} \vec{H}^{\mu\nu} \cdot \vec{H}_{\mu\nu} + m_{\rho}^{2} \:
 \vec{\rho}^{\mu} \cdot \vec{\rho}_{\mu}
 \Big)
\end{eqnarray}
in (\ref{eq:L}) describes the free mesons and
\begin{equation}
  \mathcal{L}_{\gamma} =  - \frac{1}{4} F^{\mu\nu} F_{\mu\nu}
\end{equation}
with $F^{\mu\nu}  =  \partial^{\mu} A^{\nu} - \partial^{\nu} A^{\mu}$
is the contribution of the electromagnetic field $A^{\mu}$.

The couplings $\tilde{\Gamma}_{j}$ of the mesons
$j=\sigma$, $\omega$, $\rho$ or $\delta$ 
in (\ref{eq:D}) and (\ref{eq:M}) depend
either of the vector density
\begin{equation}
  \label{eq:rho_v}
  \varrho_{v} = \sqrt{j_{\mu}j^{\mu}}
\end{equation}
with the total nucleon current 
\begin{equation}
  \label{eq:curr}
  j^{\mu} = \sum_{i=p.n} \sum_{k} w_{ik} \: \overline{\Psi}_{ik} \gamma^{\mu} \Psi_{ik}
\end{equation}
or the scalar density
\begin{equation}
  \label{eq:rho_s}
  \varrho_{s} = \sum_{i=p,n} \sum_{k} w_{ik} \: \overline{\Psi}_{ik}  \Psi_{ik} \: .
\end{equation}
where $k$ defines the single-particle state and
$w_{ik}$ is the occupation factor. 
Both quantities defined in (\ref{eq:rho_v}) and (\ref{eq:rho_s})
are Lorentz scalars.

From the Lagrangian density (\ref{eq:L}) the field equations
of all degrees of freedom are found from the Euler-Lagrange
equations treating the mesons and the electromagnetic field
as classical fields. Applying the usual mean-field and no-sea approximation
and exploiting the symmetries of a stationary system, the field equations
assume a simple form in a particular frame of reference. The
couplings $\tilde{\Gamma}_{j}$ become simple functions $\Gamma_{j}$
depending on the total vector density $n^{(v)}=n_{p}^{(v)}+n_{n}^{(v)}$
or the total scalar density $n^{(s)}=n_{p}^{(s)}+n_{n}^{(s)}$ with single
nucleon contributions given below.

In the case of nucleus with spherical symmetry, the wave function $\psi_{ik}$
of a nucleon $i$ in  single-particle states $k$ 
is a solution of the time-independent Dirac equation
\begin{equation}
  \label{eq:Dirac}
  \hat{H}_{i}
  \psi_{ik}(\vec{r}) = E_{ik} \psi_{ik}(\vec{r})
\end{equation}
with single-particle energy $E_{ik}$ and the Hamiltonian
\begin{equation}
  \label{eq:H}
  \hat{H}_{i} = 
  \left[ \vec{\alpha} \cdot \hat{\vec{p}} + \beta \left( m_{i} - S_{i} \right)
    + V_{i} + i \vec{\gamma} \cdot \frac{\vec{r}}{r} T_{i} \right]
\end{equation}
that contains three types of potentials.
Due to the symmetries, only a single component of the
Lorentz vector and isospin vector fields remains.
Thus the notation is simplified
to $\delta$, $\omega_{0}$, $\rho_{0}$ and $A_{0}$ without an additional index
for the isospin.

The scalar potential
\begin{equation}
\label{eq:S}
 S_{i}  =  g_{i\sigma} \Gamma_{\sigma} \sigma
 + g_{i\delta}\Gamma_{\delta} \delta + S^{(R)}
\end{equation}
and the vector potential
\begin{equation}
\label{eq:V}
 V_{i}  =  g_{i\omega}\Gamma_{\omega} \omega_{0}
 + g_{i\rho}\Gamma_{\rho} \rho_{0} 
 + g_{i\gamma}\Gamma_{\gamma} A_{0} + V^{(R)}
\end{equation}
contain rearrangement contributions
\begin{eqnarray}
  \label{eq:SR}
  S^{(R)} & = & 
  \frac{d\Gamma_{\sigma}}{d n^{(s)}} n_{\sigma} \sigma
  + \frac{d \Gamma_{\delta}}{d n^{(s)}} n_{\delta} \delta
  \\ \nonumber & & 
  - \frac{d \Gamma_{\omega}}{d n^{(s)}}  n_{\omega} \omega_{0}
  - \frac{d \Gamma_{\rho}}{d n^{(s)}} n_{\rho} \rho_{0}
\end{eqnarray}
if the couplings depend on the scalar density or
\begin{eqnarray}
  \label{eq:VR}
  V^{(R)} & = &
  \frac{d \Gamma_{\omega}}{d n^{(v)}} n_{\omega} \omega_{0}
 + \frac{d \Gamma_{\rho}}{d n^{(v)}}  n_{\rho} \rho_{0}
 \\ \nonumber & & 
 -  \frac{d \Gamma_{\sigma}}{d n^{(v)}} n_{\sigma} \sigma
 -  \frac{d \Gamma_{\delta}}{d n^{(v)}} n_{\delta} \delta
\end{eqnarray}
if the couplings depend on the vector density.
The derivation of these rearrangement contributions
is given in detail in appendix A.
The factors $g_{n\sigma}=g_{p\sigma}=g_{n\omega}=g_{p\omega}=
-g_{n\rho}=g_{p\rho}=-g_{n\delta}=g_{p\delta}=g_{p\gamma}=1$ and $g_{n\gamma}=0$
in (\ref{eq:S}) and (\ref{eq:V}) reflect the different coupling 
of neutrons and protons to the meson fields. The quantities $n_{\sigma}$,
$n_{\delta}$, $n_{\omega}$, and $n_{\rho}$ are source densities in
the field equations of the mesons. They are given in subsection \ref{sec:Nwfd}.
The tensor potential
\begin{equation}
  \label{eq:T}
  T_{i} = - f_{i\omega}\frac{\Gamma_{T\omega}}{m_{p}} \frac{\vec{r}}{r}
  \cdot \vec{\nabla} \omega_{0}
  - f_{i\rho}\frac{\Gamma_{T\rho}}{m_{p}} \frac{\vec{r}}{r}
  \cdot \vec{\nabla} \rho_{0}
\end{equation}
with $f_{p\omega}=f_{n\omega}=f_{p\rho}=-f_{n\rho}=1$ depends
on the derivatives of the meson fields and is particularly large
at the surface of a nucleus due to the rapid change of the meson fields.

The field equations of the mesons can be written as
\begin{eqnarray}
  \label{eq:sigma}
  -\Delta \sigma + m_{\sigma}^{2} \sigma
  & = &  \Gamma_{\sigma} n_{\sigma}
  \\
  -\Delta \delta + m_{\delta}^{2} \delta
  & = &  \Gamma_{\delta} n_{\delta}
  \\
  \label{eq:omega}
  -\Delta \omega_{0} + m_{\omega}^{2} \omega_{0}
  & = &  \Gamma_{\omega} n_{\omega}
  + \frac{\Gamma_{T\omega}}{m_{p}} \vec{\nabla} \cdot \vec{j}_{T\omega}
  \\
  \label{eq:rho}
  -\Delta \rho + m_{\rho}^{2} \rho
  & = &  \Gamma_{\rho} n_{\rho}
  + \frac{\Gamma_{T\rho}}{m_{p}} \vec{\nabla} \cdot \vec{j}_{T\rho}
\end{eqnarray}
with tensor currents
\begin{equation}
   \label{eq:tcurrent_omega}
  \vec{j}_{T\omega} = \sum_{i=p,n} \sum_{k} w_{ik} \:
  \overline{\Psi}_{ik} i \vec{\alpha} \Psi_{ik}
\end{equation}
and
\begin{equation}
  \label{eq:tcurrent_rho}
  \vec{j}_{T\rho} = \sum_{i=p,n} \sum_{k} w_{ik} \:
  \overline{\Psi}_{ik} i \vec{\alpha} \tau_{0}\Psi_{ik}
\end{equation}
for the Lorentz vector mesons.
The electromagnetic field is determined by the Poisson equation
\begin{equation}
  \label{eq:gamma}
    -\Delta A_{0} =  \Gamma_{\gamma} n_{\gamma}
\end{equation}
without a mass term.

In case of nuclear matter, the tensor potential (\ref{eq:T}) is zero and the
meson field equations reduce to a simple form
without derivative terms. Furthermore the electromagnetic
field $A_{0}$ vanishes.

\subsection{Nucleon wave functions and densities}
\label{sec:Nwfd}

In a spherical nucleus, the wave function of a nucleon
in the single-particle state $k$
is conveniently represented by the two component form
\begin{equation}
 \Psi_{ik}(\vec{r}) = \frac{1}{r} \left( \begin{array}{l}
 F_{i\kappa_{k}}(r) \mathcal{Y}_{\kappa_{k}m_{k}}(\hat{r}) \\
 i G_{i\kappa_{k}}(r) \mathcal{Y}_{-\kappa_{k}m_{k}}(\hat{r})
   \end{array}\right)
\end{equation}
with real radial wave functions $F_{i\kappa_{k}}$ and $G_{i\kappa_{k}}$.
The spin-spherical harmonics $\mathcal{Y}_{\kappa_{k}m_{k}}$ describe
the angular and spin dependence of the state that is
characterized by the 
quantum number $\kappa_{k} = \pm 1$, $\pm 2$, \dots
and the projection 
$m_{k}=-j_{k}, \dots, j_{k}$ with total angular momentum
$j_{k} = |\kappa_{k}|-1/2$ and orbital angular momentum
$l_{k} = j_{k} - \kappa_{k}/(2\left|\kappa_{k}\right|)$.

The radial wave functions $F_{i\kappa_{k}}$ and $G_{i\kappa_{k}}$
are found by solving the
eigenvalue problem
\begin{equation}
   H_{ik} 
  \left( \begin{array}{c}
    F_{i\kappa_{k}} \\ G_{i\kappa_{k}}
  \end{array} \right)
  = E_{ik}  \left( \begin{array}{c}
    F_{i\kappa_{k}} \\ G_{i\kappa_{k}}
  \end{array} \right)
\end{equation}
with the Hermitian matrix
\begin{equation}
  H_{ik} =   \left( \begin{array}{ccc}
    m_{i}-S_{i}+V_{i} & \hspace{1ex} &
    -\frac{d}{dr} -\frac{\kappa_{k}}{r} + T_{i}\\
    \frac{d}{dr} -\frac{\kappa_{k}}{r} + T_{i} & \hspace{1ex} &
    -m_{i}+S_{i}+V_{i} 
    \end{array} \right)
\end{equation}
and the boundary conditions
$F_{i\kappa_{k}}(0)=G_{i\kappa_{k}}(0)=0$ and
$\lim_{r \to 0}F_{i\kappa_{k}}(r) = \lim_{r \to 0}G_{i\kappa_{k}}(r) = 0$
for bound single-particle states. In the actual numerical calculation
the Lagrange-mesh method \cite{Typel:2018tvg} is used.

Assuming equal occupation of the sub-states for a given $\kappa_{k}$
to guarantee the sphericity of the source densities and potentials,
the single-particle vector density is given by
\begin{equation}
 n_{ik}^{(v)}  =  \frac{1}{4\pi r^{2}}
 \left[ \left| F_{ik}(r)\right|^{2} + \left| G_{ik}(r)\right|^{2} \right]
\end{equation}
with the normalization
\begin{equation}
 \int d^{3}r \: n_{ik}^{(v)}  = 1 \: .
\end{equation}
The scalar density has the form
\begin{equation}
 \qquad \mbox{and} \qquad
  n_{ik}^{(s)}  =  \frac{1}{4\pi r^{2}}
 \left[ \left| F_{ik}(r)\right|^{2} - \left| G_{ik}(r)\right|^{2} \right] 
\end{equation} 
and the tensor current can be written as
\begin{equation}
  \vec{j}_{ik}^{(t)} = n_{ik}^{(t)} \frac{\vec{r}}{r}
\end{equation}
with the density
\begin{equation}
 n_{ik}^{(t)}  =  \frac{1}{4\pi r^{2}}
 \left[ F_{ik}^{\ast}(r) G_{ik}(r) + F_{ik}(r) G_{ik}^{\ast}(r) \right] \: .
\end{equation}
A summation over all single particle states gives the total
scalar ($a=s$), vector ($a=v$) and
tensor ($a=t$) densities
\begin{equation}
  n_{i}^{(a)}  =   \sum_{k} w_{ik} n_{ik}^{(a)} \\
\end{equation}
for protons and neutrons.
The quantities $w_{ik}=0$ or $w_{ik}=1$ are the occupation factors
for the different single-particle states with $\sum_{k}w_{pk}=Z$ and
$\sum_{k}w_{nk}=N$ when $Z$ and $N$ are the charge and
neutron number of the nucleus, respectively.
Then the source terms in the meson and electromagnetic
field equations (\ref{eq:sigma}) - (\ref{eq:gamma}) can be expressed as
\begin{equation}
  \label{eq:nv}
  n_{j}  =  \sum_{i=p,n} g_{ij}  n_{i}^{(s)}
\end{equation}
for the scalar mesons $j=\sigma$, $\delta$ and
\begin{equation}
  \label{eq:ns}
  n_{j}  =  \sum_{i=p,n} g_{ij} n_{i}^{(v)}
\end{equation}
for the vector mesons $j=\omega$, $\rho$ and the
electromagnetic field $j=\gamma$. The tensor currents in
(\ref{eq:omega}) and (\ref{eq:rho}) are
\begin{equation}
  \vec{j}_{Tj}  =  \sum_{i=p,n} g_{ij} n_{i}^{(t)} \frac{\vec{r}}{r}
\end{equation}
for $j=\omega$, $\rho$. 

For nuclear matter at zero temperature the solutions of the Dirac equation
(\ref{eq:Dirac}) are simple plane waves depending on a momentum $\vec{p}_{i}$.
The summation over individual states $k$ is replaced in the
continuum approximation by an integration over momenta up to the Fermi
momentum $p_{i}^{\ast}$ and the total vector density can be written as
\begin{equation}
  \label{eq:nvi}
  n_{i}^{(v)} = g_{i} \int_{0}^{p_{i}^{\ast}} \frac{d^{3}p_{i}}{(2\pi)^{3}} =
  \frac{g_{i}}{6\pi^{2}} \left( p_{i}^{\ast}\right)^{3}
\end{equation}
with degeneracy factor $g_{i}=2$ for the spin $1/2$ nucleons.
The total scalar density is
\begin{eqnarray}
  \label{eq:nsi}
  n_{i}^{(s)} & = & g_{i} \int_{0}^{p_{i}^{\ast}} \frac{d^{3}p_{i}}{(2\pi)^{3}}
  \: \frac{m_{i}^{\ast}}{\sqrt{\left(p_{i}\right)^{2}+\left(m_{i}^{\ast}\right)^{2}}}
  \\ \nonumber & = & 
  \frac{g_{i}m_{i}^{\ast}}{4\pi^{2}} \left[ p_{i}^{\ast}\mu_{i}^{\ast}
    - \left( m_{i}^{\ast} \right)^{2}
    \ln \frac{p_{i}^{\ast}+\mu_{i}^{\ast}}{m_{i}^{\ast}} \right]
\end{eqnarray}
with the effective chemical potential
\begin{equation}
  \label{eq:mu}
  \mu_{i}^{\ast} = \mu_{i}-V_{i} = \sqrt{\left( p_{i}^{\ast}\right)^{2}
  + \left( m_{i}^{\ast}\right)^{2}}
\end{equation}
and the Dirac effective mass
\begin{equation}
  m_{i}^{\ast} = m_{i}-S_{i} \: .
\end{equation}
The tensor currents (\ref{eq:tcurrent_omega})
and (\ref{eq:tcurrent_rho}) entering the meson field equations
(\ref{eq:omega}) and (\ref{eq:rho}) give no contribution in nuclear matter
since they do not vary in space.

\subsection{Energy density functional}
\label{sec:edf}

The energy density $\varepsilon$ of the considered system is obtained
from the energy-momentum tensor. It can be expressed as a sum
\begin{equation}
  \varepsilon = \varepsilon_{\rm nucleon} + \varepsilon_{\rm field}
\end{equation}
of two contributions for a nucleus with a simple summation
\begin{equation}
 \varepsilon_{\rm nucleon} = \sum_{i=p,n} \sum_{k}  w_{ik} E_{ik} n^{(v)}_{ik}
\end{equation}
over the single-particle states in the first term and the field energy density
\begin{eqnarray}
 \varepsilon_{\rm field} & = &
 \frac{1}{2} \left( 
 \vec{\nabla} \sigma \cdot \vec{\nabla} \sigma + m_{\sigma}^{2} \sigma^{2}
 + \vec{\nabla} \delta \cdot \vec{\nabla} \delta + m_{\delta}^{2} \delta^{2}
  \right. \\ \nonumber & & \left.
  - \vec{\nabla} \omega_{0} \cdot \vec{\nabla} \omega_{0}
  - m_{\omega}^{2} \omega_{0}^{2}
  - \vec{\nabla} \rho_{0} \cdot \vec{\nabla} \rho_{0} - m_{\rho}^{2} \rho_{0}^{2}
   \right. \\ \nonumber & & \left.
  - \vec{\nabla} A_{0} \cdot \vec{\nabla} A_{0}
 \right)
 - V^{(R)} n^{(v)} + S^{(R)} n^{(s)} 
\end{eqnarray}
with rearrangement contributions and total vector and scalar densities
$n^{(v)} = n_{p}^{(v)}+n_{n}^{(v)}$ and $n^{(s)} = n_{p}^{(s)}+n_{n}^{(s)}$,
respectively. Integrating over all space the total energy
\begin{eqnarray}
  \label{eq:E_nuc}
  E & = & \sum_{i=p,n} \sum_{k} w_{ik} E_{ik}
  \\ \nonumber  & & + \frac{1}{2} \int d^{3}r \:
  \Big\{ \Gamma_{\sigma} n_{\sigma} \sigma
  + \Gamma_{\delta} n_{\delta} \delta
  - \Gamma_{\gamma} n_{\gamma} A_{0}
  \\ \nonumber & &
    - \left[ \Gamma_{\omega} n_{\omega}
    + \frac{\Gamma_{T\omega}}{m_{p}} \vec{\nabla} \cdot \vec{j}_{T\omega}
    \right] \omega_{0}
    \\ \nonumber & &
  - \left[ \Gamma_{\rho} n_{\rho}
    + \frac{\Gamma_{T\rho}}{m_{p}} \vec{\nabla} \cdot \vec{j}_{T\rho}
    \right] \rho_{0}
  \\ \nonumber & &
   - V^{(R)} n^{(v)} + S^{(R)} n^{(s)} 
  \Big\} 
\end{eqnarray}
is found with explicit tensor contributions when the field equations and partial
integrations are applied.

In the case of cold nuclear matter the energy density can be expressed as
\begin{eqnarray}
  \label{eq:eps}
  \varepsilon & = & \sum_{i=p,n} \varepsilon_{i}^{\rm kin}
  + \frac{1}{2} \sum_{j=\sigma,\delta} \left( C_{j}
  + n^{(s)} D_{j}^{(s)} \right) n_{j}^{2}
  \\ \nonumber & &
  + \frac{1}{2} \sum_{j=\omega,\rho} \left( C_{j}
  - n^{(s)} D_{j}^{(s)} \right) n_{j}^{2}
\end{eqnarray}
with the kinetic contribution
\begin{equation}
  \varepsilon_{i}^{\rm kin} = \frac{1}{4} \left[ 3\mu_{i}^{\ast}n_{i}^{(v)}
    + m_{i}^{\ast}n_{i}^{(s)} \right]
\end{equation}
and the factors
$ C_{j} = \Gamma_{j}^{2}/m_{j}^{2}$
and their derivatives
$D_{j}^{(s)} = dC_{j}/d n^{(s)}$
with respect to the scalar density.
The latter are nonzero only for couplings $\Gamma_{j}$ that depend on
the scalar density $n^{(s)}$. The pressure assumes the form
\begin{eqnarray}
  \label{eq:pres}
  p & = & \sum_{i=p,n} p_{i}^{\rm kin}
  - \frac{1}{2} \sum_{j=\sigma,\delta} \left( C_{j}
  + n^{(s)} D_{j}^{(s)} + n^{(v)} D_{j}^{(v)} \right) n_{j}^{2}
   \nonumber \\ & &
  + \frac{1}{2} \sum_{j=\omega,\rho} \left( C_{j} + n^{(s)} D_{j}^{(s)}
  + n^{(v)} D_{j}^{(v)}\right) n_{j}^{2}
\end{eqnarray}
with the kinetic contribution
\begin{equation}
  \label{eq:pkin}
  p_{i}^{\rm kin} = \frac{1}{4} \left[ \mu_{i}^{\ast}n_{i}^{(v)}
    - m_{i}^{\ast}n_{i}^{(s)} \right] 
\end{equation}
and the derivatives 
$D_{j}^{(v)} = dC_{j}/d n^{(v)}$
with respect to the vector density. In contrast to the pressure
(\ref{eq:pres}) there are no contributions from terms with
$D_{j}^{(v)}$ in the energy density (\ref{eq:eps}).
The thermodynamic relation
\begin{equation}
  \varepsilon + p = \sum_{i=p,n} \mu_{i} n_{i}^{(v)}
\end{equation}
with the chemical potentials $\mu_{i}$, see (\ref{eq:mu}), is easily verified.

For the calculation of the incompressibility $K$ of symmetric nuclear matter
at the saturation density $n_{\rm sat}$ it
is useful to know the density derivative of the pressure in isospin
symmetric nuclear matter.
For this purpose also the second derivatives
$E_{j}^{(s)} = d^{2}C_{j}/d (n^{(s)})^{2}$ and
$E_{j}^{(v)} = d^{2}C_{j}/d (n^{(v)})^{2}$ are needed. With
$n^{(s)}=n_{\sigma}$ and
$n^{(v)}=n_{\omega}$ in this case, the derivative of the pressure is given by
the lengthy general expression
\begin{eqnarray}
  \label{eq:dpdnv}
  \frac{dp}{dn^{(v)}} & = &
    \frac{1}{3} \left[
     \mu^{\ast}
    -  m^{\ast}\frac{dn^{(s)}}{dn^{(v)}} \right]
   \\ \nonumber & &
   -  \left(  D_{\sigma}^{(v)}
   + \frac{1}{2} n^{(v)} E_{\sigma}^{(v)} \right) (n^{(s)})^{2}
   \nonumber \\ & &
   +  \left( D_{\omega}^{(v)} 
   + \frac{1}{2} n^{(v)} E_{\omega}^{(v)}
   \right) (n^{(v)})^{2}
  \nonumber \\ & &
  +  \left( C_{\omega} + n^{(s)} D_{\omega}^{(s)}
  + n^{(v)} D_{\omega}^{(v)}\right) n^{(v)}
     \\ \nonumber & &
     -   \left( C_{\sigma}
   + n^{(s)} D_{\sigma}^{(s)} + n^{(v)} D_{\sigma}^{(v)} \right)
   n^{(s)} \frac{dn^{(s)}}{dn^{(v)}}
   \\ \nonumber & &
   -   \left(  D_{\sigma}^{(s)}
   + \frac{1}{2} n^{(s)} E_{\sigma}^{(s)}
    \right)  \frac{dn^{(s)}}{dn^{(v)}} (n^{(s)})^{2}
   \nonumber \\ & &
   + \left(   D_{\omega}^{(s)}
   + \frac{1}{2} n^{(s)} E_{\omega}^{(s)} 
   \right)  \frac{dn^{(s)}}{dn^{(v)}} (n^{(v)})^{2}
\end{eqnarray}
with the derivative
\begin{eqnarray}
  \label{eq:dnsdnv}
  \frac{dn^{(s)}}{dn^{(v)}} & = &
  \left[  \frac{m^{\ast}}{\mu^{\ast}} 
  +f \left( D_{\omega}^{(s)} n^{(v)} - D_{\sigma}^{(v)} n^{(s)}\right) \right]
  \\ \nonumber & & \left\{ 1
+   f\left[  C_{\sigma}
  + \frac{1}{2}E_{\sigma}^{(s)} (n^{(s)})^{2}
   \right. \right. \\ \nonumber & & \left. \left.
    + 2 D_{\sigma}^{(s)} n^{(s)}
    - \frac{1}{2}E_{\omega}^{(s)}  (n^{(v)})^{2}
    \right] \right\}^{-1}
\end{eqnarray}
that contains the factor
\begin{equation}
  \label{eq:facf}
   f = 3 \left( \frac{n^{(s)}}{m^{\ast}} - \frac{n^{(v)}}{\mu^{\ast}}\right) \: .
\end{equation}
Then the incompressibility is found from
\begin{equation}
  K 
    = 9 \left.  \frac{dp}{dn^{(v)}} \right|_{n_{\rm sat}}
\end{equation}
at the saturation density 
For a vector density dependence of the
couplings, the terms with $D_{j}^{(s)}$ and $E_{j}^{(s)}$ vanish.
Analogously, $D_{j}^{(v)}=E_{j}^{(v)}=0$ for a scalar density dependence.

The symmetry energy of nuclear matter is given by
\begin{equation}
  \label{eq:esym}
  E_{\rm sym}(n_{b}) = \frac{1}{2} \frac{\partial^{2}}{\partial \delta^{2}}
  \frac{E}{A}
\end{equation}
as a second derivative of the energy per nucleon
\begin{equation}
  \frac{E}{A} = \frac{\epsilon}{n_{b}}
\end{equation}
with respect to the neutron-proton asymmetry
$\delta = (n_{n}^{(v)}-n_{p}^{(v)})/n_{b}$
where $n_{b}=n^{(v)}=n_{n}^{(v)}+n_{p}^{(v)}$ is the baryon density.
The symmetry energy at saturation $J=E_{\rm sym}(n_{\rm sat})$
and the slope parameter $L = \left. dE_{\rm sym}/dn_{b} \right|_{n_{b}=n_{\rm sat}}$
are then easily obtained. The so-called volume part
of the isospin incompressibility \cite {Dutra:2014qga}
\begin{equation}
  K_{\tau,v} = K_{\rm sym}-6L-\frac{QL}{K} 
\end{equation}
quantifies the change of the incompressibility of nuclear matter
with the isospin asymmetry $\delta$.
It  contains the incompressibility of the symmetry energy
\begin{equation}
  K_{\rm sym} = 9 n_{\rm sat}^{2}
  \left.  \frac{d^{2} E_{\rm sym}}{d(n_{b})^{2}}  \right|_{n_{\rm sat}} \: ,
\end{equation}
the quantities $L$ and $K$, and the skewness parameter
\begin{equation}
  Q = 27 n_{\rm sat}^{3}
  \left.  \frac{\partial^{2}}{\partial(n_{b})^{3}} \frac{E}{A}
  \right|_{n_{\rm sat},\delta = 0}
\end{equation}
which is proportional to the third derivative of the energy per nucleon
with respect to the baryon density. It describes the deviation of $E/A$
in symmetric nuclear matter from
a parabolic dependence on $n_{b}$ near the saturation point.

\section{Determination of model parameters}
\label{sec:para}

The nuclear EDF with density dependent couplings
constitutes a phenomenological approach to describe properties
of nuclei and nuclear matter. It depends on some parameters
that have to be determined so that the calculated observables agree 
with experimental data as far as possible.
The statement of 'best agreement' has to be made
more precise and quantitative by defining an objective
function that depends on the
selected observables, the associated uncertainties and its functional form.
Once it is fixed the parameters of the model can be found
by appropriate numerical
fitting strategies. Usually the objective function is not changed during the
process of parameter determination. However, in the present study,
the uncertainties of the different observables will be adjusted
in order to find values that give a
true representation of their size.

\subsection{Parameters and density dependence of couplings}

Some parameters in the density functional are kept constant during the fitting
procedure. These are the masses of the nucleons and that of the
$\omega$, $\rho$, and $\delta$ meson. In this work the same values
as in \cite{Typel:2018cap} are used, i.e.,
$m_{p}=938.272081$~MeV, $m_{n}=939.565413$~MeV, $m_{\omega}=783$~MeV,
$m_{\rho}=763$~MeV,
and $m_{\delta}=980$~MeV. In constrast, the mass $m_{\sigma}$
of the $\sigma$ meson is used as
a variable parameter because it has the longest range of the mesons
with the strongest finite-range effects.

All other parameters enter via the couplings of the mesons
with the nucleons. Their number will depend on the functional form of the
density dependence and the way it is parametrised. All couplings can be written
in the form
\begin{equation}
  \Gamma_{j}(n) = \Gamma_{j}^{(0)} f_{j}(x)
\end{equation}
with a constant $\Gamma_{j}^{(0)}= \Gamma_{j}(n_{\rm ref})$ at a reference density
$n_{\rm ref}$ and an arbitrary function $f_{j}$ that depends on an argument
$x=n/n_{\rm ref}$ with condition $f_{j}(1)=1$.
The density $n$ can be the vector density $n^{(v)}$
or the scalar density $n^{(s)}$.  In the present work we use the rational form
\begin{equation}
  \label{eq:fso}
  f_{j}(x) = a_{j} \frac{1+b_{j}(x+d_{j})^{2}}{1+c_{j}(x+d_{j})^{2}}
\end{equation}
as introduced in \cite{Typel:1999yq}
with four parameters $a_{j}$, $b_{j}$, $c_{j}$, and $d_{j}$ for the isoscalar
$\sigma$ and $\omega$ meson. The normalisation condition at $n_{\rm ref}$, i.e.,
$x=1$, fixes
$a_{j}=[1+c_{j}(1+d_{j})^{2}]/[1+c_{j}(1+d_{j})^{2}]$. As a second constraint
on the function (\ref{eq:fso}) the condition $f_{j}^{\prime\prime}(0)=0$
is demanded.
This leads to the relation $3c_{j}d_{j}^{2}=1$. In total there are only two
independent parameters, e.g., $b_{j}$ and $c_{j}$,
for the density dependence of the isoscalar mesons $\omega$ and $\sigma$.
For the isovector $\rho$
and $\delta$ mesons the same exponential form
\begin{equation}
  \label{eq:frho}
  f_{j}(x) = \exp[-a_{j}(x-1)]
\end{equation}
as in \cite{Typel:1999yq}
is chosen with a single parameter $a_{j}$. 
The tensor couplings $\Gamma_{T\omega}$ and $\Gamma_{T\rho}$ are assumed to be
constant.
In principle, the parameters entering the coupling
functions can be used directly
in the fitting procedure. However, they can be highly correlated and it is more
convenient to use nuclear matter parameters as independent variables. See 
appendix B for the conversion of one set of parameters to the other.

\subsection{Observables}

The parameters of the relativistic energy density functional can be determined
by fitting experimental observables of atomic nuclei and empirical data
of nuclear matter that are obtained with extrapolation
methods from nuclear observables.
Since these nuclear matter quantities might be affected by some model
dependence in such a process,
only directly observable quantities of nuclei are considered in this work.

There are three types of nuclear observables that are considered in this work:
binding energies, quantities related to the charge form factor, and spin-orbits
splittings derived from single-particle energies of the nucleons.
The binding energy $B_{N,Z}$ of a nucleus with $N$ neutrons and $Z$ protons
is obtained from the total energy (\ref{eq:E_nuc}). Since the long-range
behaviour of the Coulomb field, which is obtained from solving
equation (\ref{eq:gamma}), is not correct for the motion of a proton
with respect to the core of a nucleus in the mean-field approximation,
a correction is applied by multiplying the field with a factor $(Z-1)/Z$
for a nucleus with $Z$ protons.
The theoretical values of the binding energies
can only be compared to experimental data if
a center-of-mass correction $E_{N,Z}^{(\rm cm)}$ is subtracted from the
total energy (\ref{eq:E_nuc}). This correction is calculated as
the expectation value
\begin{equation}
 E_{N,Z}^{(\rm cm)} = \langle \frac{\hat{\vec{P}}^{2}}{2M_{N,Z}} \rangle
\end{equation}
from the total many-body wave function
with the total momentum $\hat{\vec{P}}=\sum_{n=1}^{A} \hat{\vec{p}}_{n}$
that is as sum of all nucleon momenta and
$M_{N,Z}=Nm_{n}+Zm_{p}$. A center-of-mass correction is also applied in the
calculation of the charge form factor from the charge distribution
of the nucleus
as explained in \cite{Typel:2018cap}. From the charge form factor the charge
radius, diffraction radius and surface thickness are extracted,
see \cite{Rufa:1988zz}
for details. Finally spin-orbit splittings of nuclear level pairs with
equal $l$ and $j=l\pm 1/2$ close to the Fermi energies of neutrons and protons
are used as observables. Their values
are obtained from the excitation energy spectrum
of neighboring nuclei with one nucleon more or less than the
nucleus of interest.
This method is most reliable only for closed-shell nuclei where prominent
single-particle resonances are found and a fragmented
distribution of strength over several levels is not observed.

Since pairing effects are not included in the present density
functional calculations,
the set of nuclei in the fit of the parameters reduces further to those where
these effects are unimportant. Thus the set of nuclei used here
is rather limited. It comprises ${}^{16}$O, ${}^{24}$O, ${}^{40}$Ca,  ${}^{48}$Ca,
${}^{56}$Ni, ${}^{90}$Zr, ${}^{100}$Sn, ${}^{132}$Sn, and ${}^{208}$Pb
but the essential effects of the choice of the couplings on the
quality of the fit can already be seen clearly. The experimental
data for these nuclei are given in table~1 of reference
\cite{Typel:2018cap}. There are $N_{1}^{\rm (obs)}=9$ experimental
data for the binding energy, $N_{2}^{\rm (obs)}=6$ for the charge radius,
$N_{3}^{\rm (obs)}=N_{4}^{\rm (obs)}=5$
for the diffraction radius and the surface thickness, and
$N_{5}^{\rm (obs)}=12$ for the spin-orbits splittings,
hence $N_{\rm obs}=5$ different observables and $N_{\rm data}=37$
experimental data in total.

\subsection{Objective function and uncertainties}

The optimal fit of the model parameters to the experimental data is found
from a minimisation of the function
\begin{eqnarray}
  \label{eq:chi2}
  \chi^{2}(\{p_{k}\})
  =  \sum_{i=1}^{N_{\rm obs}}  \chi_{i}^{2}(\{p_{k}\})
\end{eqnarray}
with
\begin{eqnarray}
  \label{eq:chi2i}
  \chi_{i}^{2}(\{p_{k}\})
  =  \sum_{n=1}^{N_{i}^{\rm (obs)}}
  \left[ \frac{O_{i}^{\rm (exp)}(n)
      -O_{i}^{\rm (model)}(n,\{p_{k}\})}{\Delta O_{i}} \right]^{2}
\end{eqnarray}
for each observable $i$ by varying the parameters $p_{k}$.
This is achieved in the multidimensional parameter space
by applying the simplex method
as specified in \cite{NumericalRecipes}.
However, the uncertainties $\Delta O_{i}$
are not kept constant during the fit as in \cite{Typel:2018cap}.
Before the iteration starts, reasonable values of the parameters $p_{k}$
and their possible variation $\Delta p_{k} \approx p_{k}/100$ are defined.
Also appropriate uncertainties $\Delta O_{i}$ are chosen guided by typical
uncertainties of EDFs, e.g., $1.5$~MeV for binding energies, $0.5$~MeV for
spin-orbit splittings and $0.02$~fm for radii and surface thicknesses.
During the iteration, the corner points of the simplex are moved in
the direction of lower $\chi^{2}$ and the allowed variation $\Delta p_{k}$
of the parameters
is adjusted depending on the needed expansion or contraction of the simplex
to reduce $\chi^{2}$.
After one hundred iterations through all parameters
the uncertainties of the observables $\Delta O_{i}$ are re-scaled
so that
$\chi_{i}^{2}(\{p_{k}\})/N_{i}^{\rm (obs)}$ is the same for all observables and
\begin{equation}
  \chi^{2}(\{p_{k}\})/N_{\rm dof}=1
\end{equation}
with the number of degrees
of freedom $N_{\rm dof} = N_{\rm data} - N_{\rm par}$ and the number of data
$ N_{\rm data} = \sum_{i=1}^{N_{\rm obs}} N_{i}^{\rm (obs)}=37$. This
means that every experimental data point
contributes on the average equally to the
total uncertainty. The cycle of iterations is repeated several times until
the variation of the parameters falls below $2\cdot 10^{-6}$. The
procedure is also repeated with different initial conditions to check
for the stability of the obtained minimum.
Under this condition,
the obtained uncertainties are determined self-consistently
with the energy density
functional and have a reasonable size. Thus more information is obtained than
in a simple $\chi^{2}$ fit with uncertainties that are fixed from the outset.

From the $\chi^{2}$ function (\ref{eq:chi2}) the symmetric matrix
\begin{equation}
  \mathcal{M}_{ij} = \frac{1}{2}
  \left. \frac{\partial^{2} \chi^{2}}{\partial p_{i} \partial p_{j}} \right|_{\vec{p}^{\rm min}}
\end{equation}
can be formed at the position $\vec{p}^{\rm min}=(p_{1}^{\rm min}, \dots ,
p_{N_{\rm obs}}^{\rm min})$ of minimum $\chi^{2}$.
It is used to calculate the covariance
\begin{equation}
  \overline{\Delta A \Delta B}
  = \sum_{ij} \frac{\partial A}{\partial p_{i}}
  \left( \mathcal{M}^{-1}\right)_{ij} \frac{\partial B}{\partial p_{j}}
\end{equation}
of any two observables $A$ and $B$
\cite{Reinhard:2010wz}. Then the uncertainty (one-$\sigma$ confidence level)
of an observable $A$ can be defined as
\begin{equation}
  \Delta A = \sqrt{\overline{\left(\Delta A\right)^{2}}}
\end{equation}
and the correlation coefficient is given by 
\begin{equation}
  c_{AB} = \frac{\overline{\Delta A \Delta B}}{\sqrt{
      \overline{\left(\Delta A\right)^{2}} \:
      \overline{\left(\Delta B\right)^{2}}}}
\end{equation}
that assumes values between $-1$ and $1$. At these limits the two observables
are fully (anti)correlated. The case $c_{AB}=0$ means that the observables
are totally uncorrelated.

\subsection{Selection of energy density functionals}

The effect of the tensor couplings on the quality of the theoretical
description of nuclei is most easily seen in comparison to conventional
relativistic EDFs with density dependent couplings
that are obtained with the same strategy to fit the
model parameters. Hence, a variety of models is considered in this work.
The basic EDF contains only $\omega$, $\sigma$, and $\rho$ mesons that
couple minimally to the nucleons. This type of model contains eight independent
parameters that are found with the methods as described above. Then
the EDF is extended to include also $\omega$ and $\rho$ meson tensor
couplings increasing the number of parameters to ten. In a further step,
the $\delta$ meson is included as a new degree of freedom with
one additional parameter. The nuclear incompressibility is kept
fixed in all these EDFs at $K=240$~MeV, a representative value
of relativistic mean-field models \cite {Dutra:2014qga} that
is inside the range of favoured values from the analysis of the energy
of isoscalar giant monopole resonances in nuclei, however, no unanimous
agreement has been reached so far, see, e.g., \cite{Stone:2014wza}.
Without using this observable directly in the fit
of parameters it is difficult to obtain reasonable values of $K$.
For these three types of EDFs one set with couplings depending on the
vector density and one set with scalar density dependencies are investigated
bringing the total number of models to six. 

\section{Results}
\label{sec:res}

\subsection{Couplings}

After performing the fit of the EDFs to the properties of finite
nuclei using the approach as described in section \ref{sec:para},
the parameters for the best description are obtained. The numerical
values are given in tables \ref{tab:01} and \ref{tab:02}, including
the mass of the $\sigma$ meson, the couplings at the reference density
(vector or scalar) and
the parameters defining the density dependence of the couplings.

Some obvious correlations
of individual quantities with the type of EDF are found.
The introduction of tensor couplings (models DDVT, DDST) leads
to reductions of the $\sigma$ meson mass and of the $\omega$ and $\rho$
coupling strengths as compared to the standard models (DDV, DDS).
This feature is related to the increased Dirac effective mass, see below.
The ratio $\Gamma_{\sigma}/m_{\sigma}$, which is the relevant quantity
for calculations of nuclear matter, changes less strongly between the models.
The $\rho$ meson tensor coupling is substantially larger than the
$\omega$ meson tensor coupling as observed, e.g., already in
\cite{Reinhard:1989zi,Bender:1999yt}. 
Also an increase of the reference densities,
$n_{\rm ref}^{(v)}$ or $n_{\rm ref}^{(s)}$,
is seen. The further introduction of the $\delta$ meson (DDVTD, DDSTD)
only leads to small
changes of the parameters, with the exception of the $\rho$ meson
coupling that becomes larger.
For models with scalar
density dependence and tensor coupling, there are two unique cases
(DDST, DDSTD) where
the parameters in the function (\ref{eq:fso}) become very small ($c_{\omega}$)
or even negative ($b_{\omega}$), see table \ref{tab:02}.
The latter case would cause the coupling to vanish
and to become negative at very high densities. However, this is not relevant
for calculations of nuclear structure or nuclear matter at reasonable
baryon densities since they are much lower than the zero-crossing densities.

\begin{figure}
\resizebox{0.48\textwidth}{!}{%
  \includegraphics{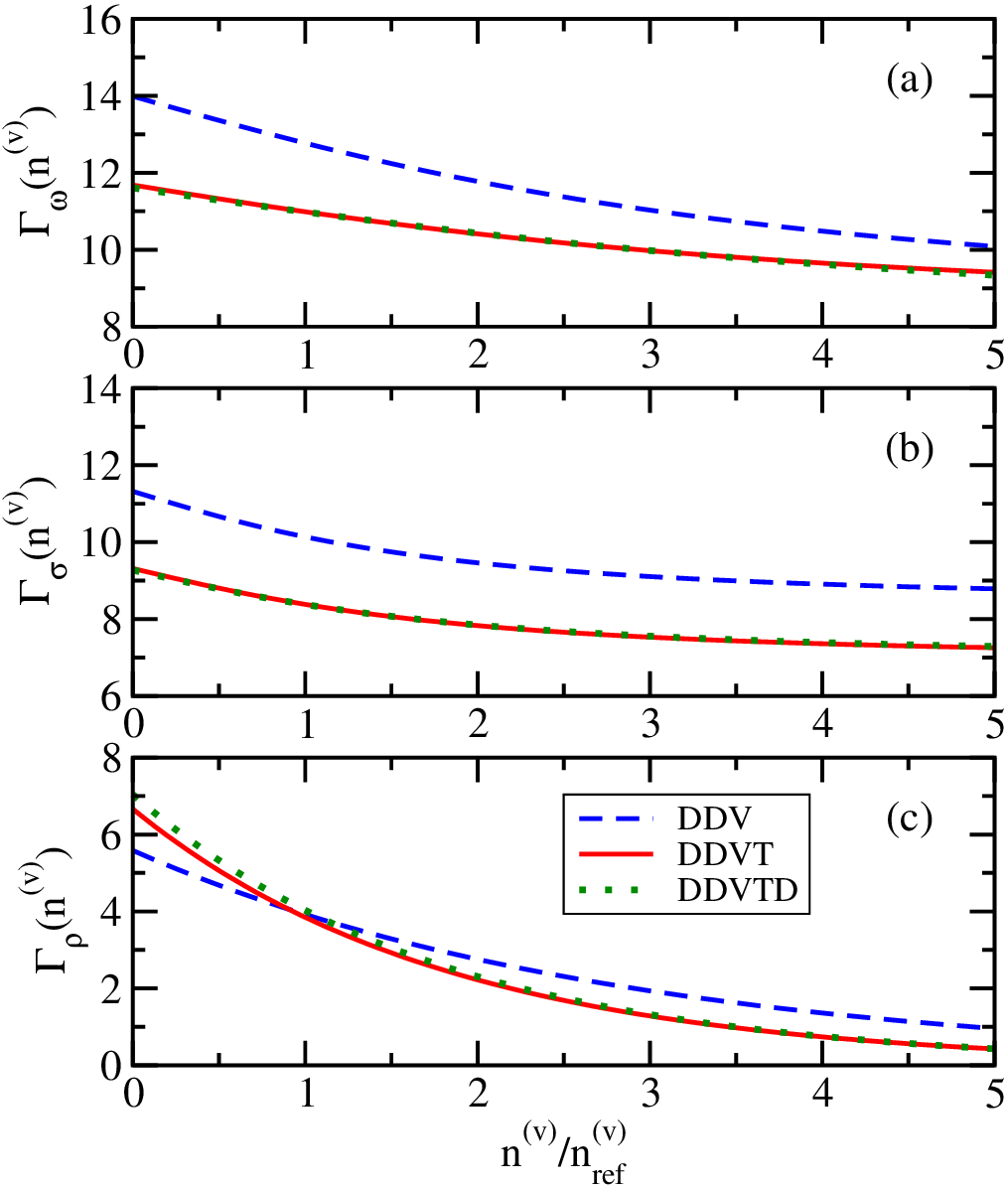}
}
\caption{Coupling functions of the $\omega$ (a), $\sigma$ (b), and
$\rho$ (c) meson for models with a vector density dependence.}
\label{fig:01}
\end{figure}

\begin{figure}
\resizebox{0.48\textwidth}{!}{%
  \includegraphics{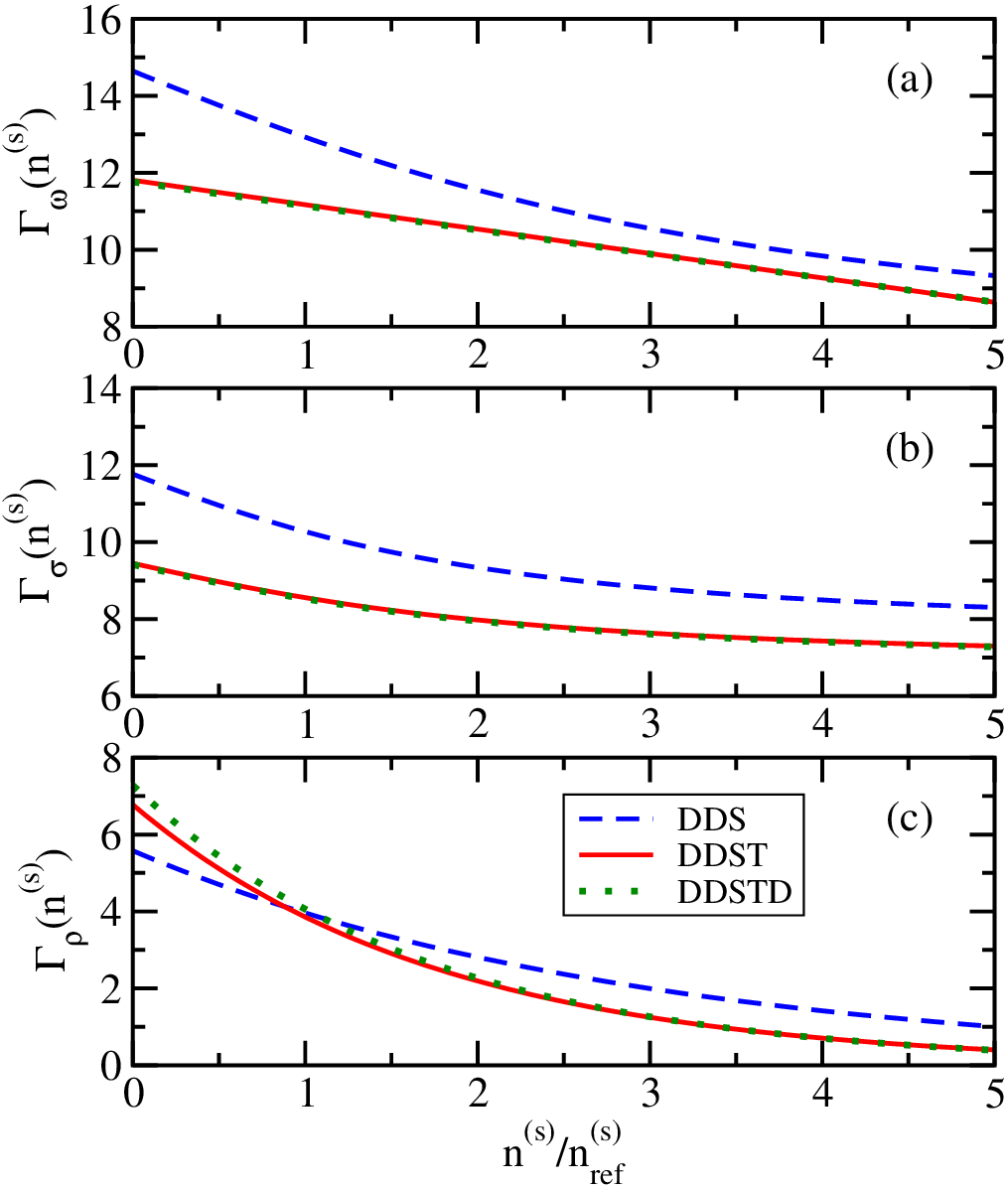}
}
\caption{Coupling functions of the $\omega$ (a), $\sigma$ (b), and
$\rho$ (c) meson for models with a scalar density dependence.}
\label{fig:02}
\end{figure}

The actual density dependence of the couplings is depicted
in figures \ref{fig:01} and \ref{fig:02} for the cases of a vector or scalar
density dependence, respectively.
Only the $\omega$, $\sigma$ and $\rho$ couplings
are shown because the $\delta$ coupling has the same shape as the $\rho$
coupling if it is nonzero. A typical decrease of the couplings with
increasing density is observed. 
All couplings behave rather similarly. The $\rho$ meson coupling decreases more
strongly than the $\omega$ and $\sigma$ couplings. It vanishes
at infinitely high density because the exponential form (\ref{eq:frho})
was chosen. The situation is different for the isoscalar mesons. They approach
a nonzero finite value in this limit.
The variations between the parametrisations
are less strong for the $\rho$ meson as compared to the isoscalar mesons.

\subsection{Uncertainties of observables}

\begin{figure}
\resizebox{0.48\textwidth}{!}{%
  \includegraphics{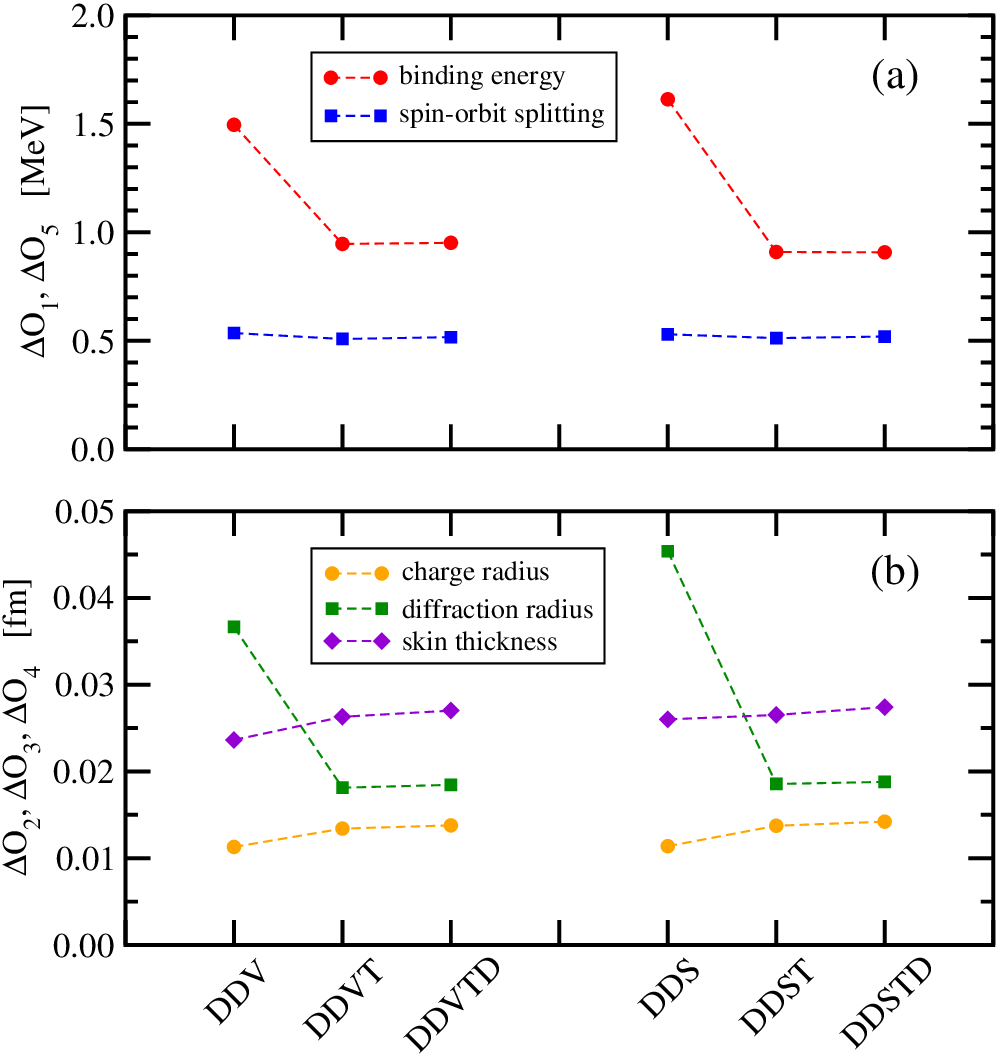}
}
\caption{Uncertainties $\Delta O_{i}$ of the parametrisation in
  energy observables (a) and length observables (b).}
\label{fig:03}
\end{figure}

The introduction of tensor couplings in the energy density functional also
affects the uncertainties of nuclear observables that enter in the calculation
of the $\chi^{2}$ function (\ref{eq:chi2i}).
They are given in table \ref{tab:03}
and shown in figure \ref{fig:03}. Most striking is the reduction of the
uncertainty in the binding energies (upper panel) and in the diffraction radii
(lower panel) when the tensor couplings are considered.
In contrast, the charge radii and skin thicknesses are only
described slightly worse than in the models without tensor interaction.
Taking the $\delta$ meson into account does not make a big difference.
The uncertainties of the
spin-orbit splittings are almost the same for all models.
The observed trends are very similar
for models with a vector or a scalar density dependence of the couplings.
Overall, terms with tensor couplings seem to be a valuable contribution
in the EDF to improve the description of nuclear observables.

\subsection{Nuclear matter parameters}

With the fitted parameters of the energy density functionals, the
characteristic parameters of nuclear matter can be calculated easily
from the dependence of the energy density (\ref{eq:eps}) on the baryon density
$n_{b}$ and the isospin asymmetry $\delta$. They are given in tables
\ref{tab:04} and \ref{tab:05} with uncertainties, including the
$\sigma$ meson mass and the average Dirac effective mass
$m^{\ast}=(m_{n}^{\ast}+m_{p}^{\ast})/2$ at saturation in symmetric nuclear matter.

The most notable result of including tensor couplings is the substantial
rise of the Dirac effective mass at saturation and the drop in the mass
of the $\sigma$ meson. At the same time the uncertainties of these
two quantities rise in the parametrisations with tensor couplings.
In conventional relativistic
energy density functionals without tensor couplings the Dirac effective mass
has to be rather small in order to find a reasonable size of spin-orbit
splittings in nuclei. The proper binding energy per nucleon is primarily
determined by the difference $V-S$ of the vector and scalar potentials whereas
the spin-orbit splitting depends on the gradient of the sum $V+S$. Hence, in
order to describe both observables reasonably well, strong constraints are
put on the size of $V$ and $S$ themselves and thus on the couplings
of the $\omega$ and $\sigma$ mesons. With tensor couplings, there is an
additional contribution to the spin-orbit potential. The Dirac effective mass
can be larger and the $\sigma$ meson coupling smaller than usual, cf.\
table \ref{tab:01}. Since effects of the tensor couplings are related to the
variation of the densities, it is of no surprise that other quantities that are
sensitive to the description of the nuclear
surface are also affected. This is particular true
for the mass of the $\sigma$ meson because its low mass determines the range of
the interaction. 

Another effect of tensor couplings is the slight increase of the saturation
density of nuclear matter, whereas the binding energy per nucleon is less
affected as seen in table \ref{tab:04}. These two quantities are rather well
constrained in the fit as the small uncertainties show.
The skewness $Q$, however, 
is varying strongly between the models and is rather unconstrained, even though
large negative values are preferred.

The symmetry energy $J$ and its slope parameter $L$ are also influenced
by the introduction of the tensor couplings as seen from table \ref{tab:05}.
Both values reduce, in particular the slope parameter, when tensor couplings are
introduced. The uncertainty of $J$ is usually in the order of $7$\% for
the DDV and DDS models or below $4.5$\% for the other parametrisations.
The slope parameter $L$ is much less constrained with uncertainties around
$30$\%. The symmetry energy incompressibilities
$K_{\rm sym}$ are always found to be negative with values correlated to
$J$ and $L$. The (negative) values of $K_{\tau,v}$ vary over
a much larger range. In both cases the uncertainties are much larger
for parametrisations with a scalar density dependence of the couplings.

\subsection{Equation of state and symmetry energy}

\begin{figure}
\resizebox{0.48\textwidth}{!}{%
  \includegraphics{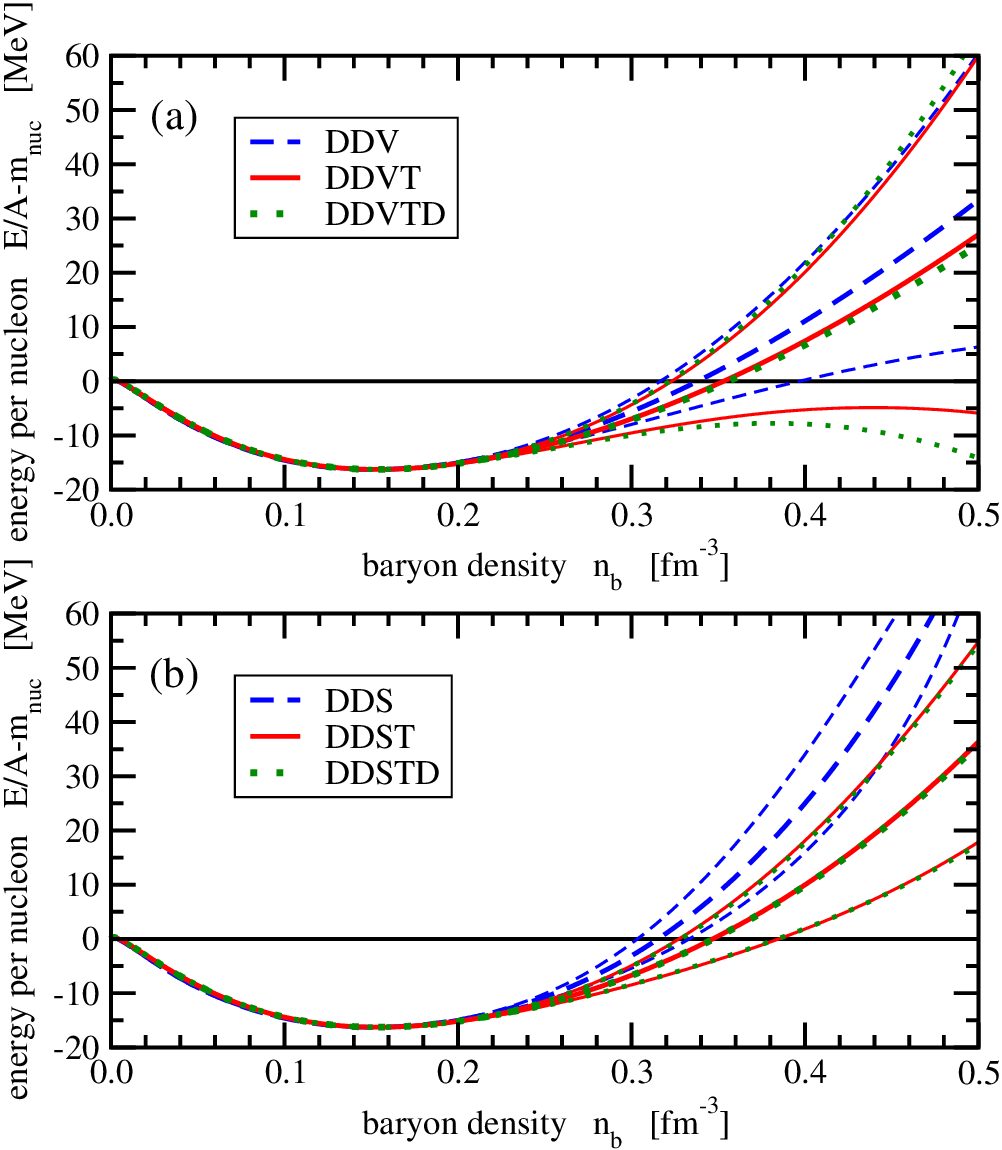}
}
\caption{Equation of state of symmetric nuclear matter with
 coupling functions depending on the vector density (a) and on
 the scalar density (b). See text for details.}
\label{fig:04}
\end{figure}

\begin{figure}
\resizebox{0.48\textwidth}{!}{%
  \includegraphics{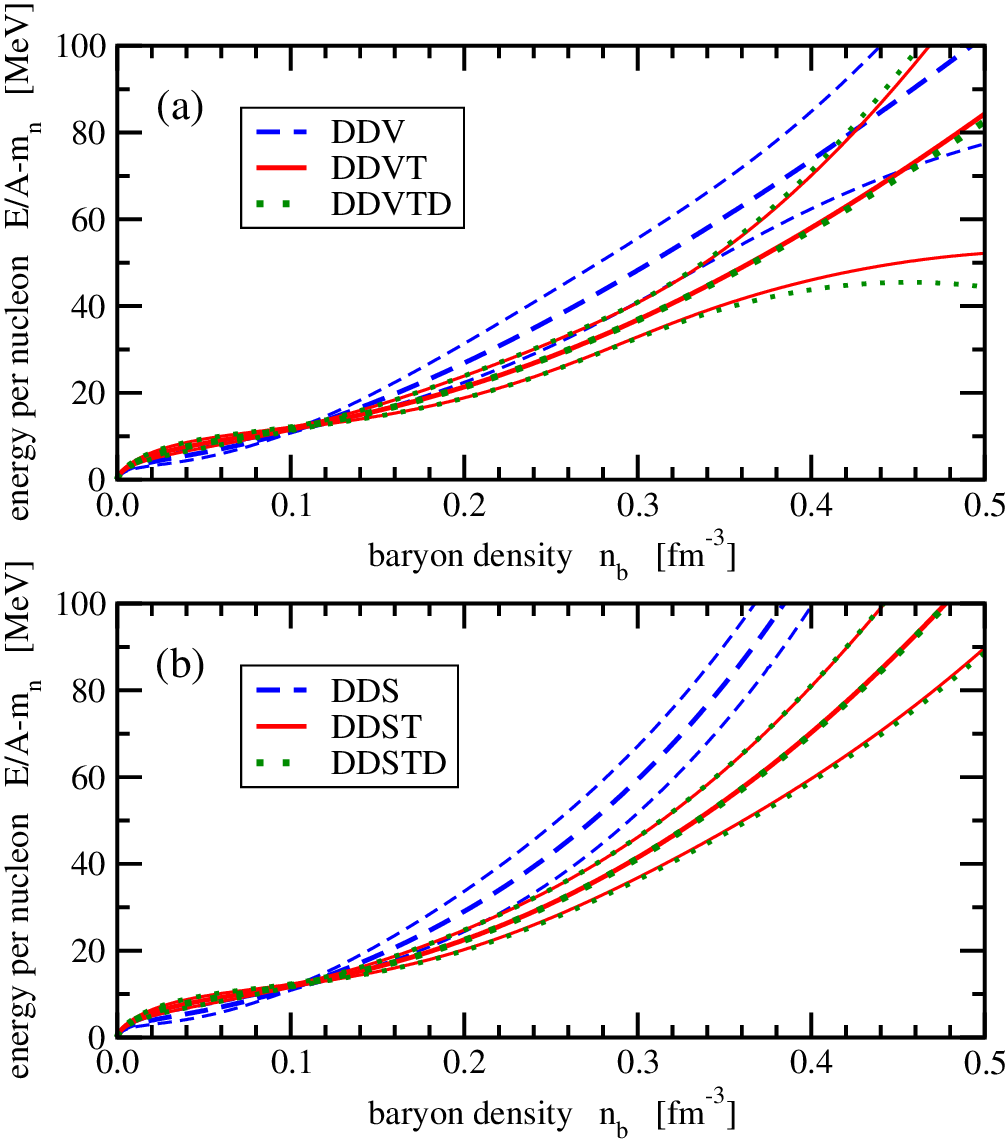}
}
\caption{Equation of state of pure neutron matter with
  coupling functions depending on the vector density (a) and on
  the scalar density (b). See text for details.}
\label{fig:05}
\end{figure}

The nuclear matter parameters characterise the equation of state
(EoS) only close
to saturation density and isospin asymmetry zero. 
A larger range of baryon densities is explored in figures \ref{fig:04}
and \ref{fig:05} for symmetric nuclear matter and pure neutron matter,
respectively. The results of the calculations with the best parameters are
depicted with thick lines or dots. The uncertainty band is given by the
corresponding thin lines and small dots with the same colors.
There is hardly any difference between the models for the
energy per nucleon up to nuclear saturation density. Different trends
are seen at higher densities. This is obvious because
the fit of the models to observables of nuclei is sensitive
essentially only to the sub-saturation region as
higher densities are not found in finite nuclei.

The models with a vector density dependence of the couplings (upper panel of
figure \ref{fig:04}) behave very similarly in the case of symmetric
nuclear matter with a slight reduction of the stiffness when tensor couplings
are included in the model.
The ordering is similar in case of pure neutron matter (upper panel
of figure \ref{fig:05}) but the effect of softening is a bit stronger.

The situation is different for models with a scalar density dependence.
Here, the models DDST and DDSTD are almost identical and much softer than
the other two parametrisations. This is true for symmetric nuclear matter
as well as pure neutron matter.
The EoS of model DDS is stiffer at high densities than that of DDV,
both without tensor interaction. A similar trend is observed for models
with tensor interaction. This feature will be explained below when the
evolution of effective masses and scalar densities is studied.

\begin{figure}
\resizebox{0.48\textwidth}{!}{%
  \includegraphics{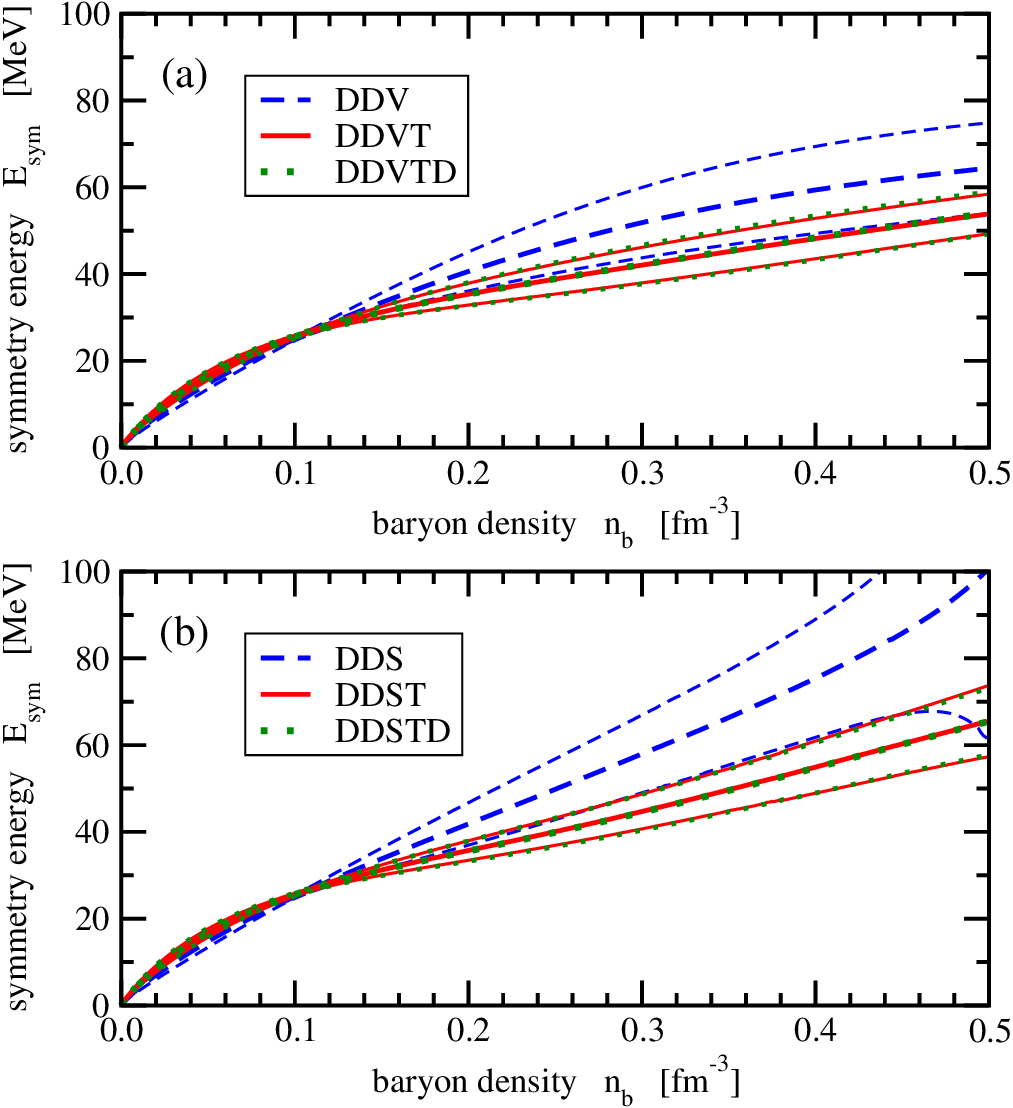}
}
\caption{Symmetry energy as a function of the baryon density with
  coupling functions depending on the vector density (a) and on
  the scalar density (b). See text for details.}
\label{fig:06}
\end{figure}

The density dependence of the nuclear symmetry calculated according to
(\ref{eq:esym}) is shown 
in figure \ref{fig:06}, again for models with vector density dependence
(upper panel) and scalar density dependence (lower panel). For baryon densities
below saturation there is no big difference between the parametrisations
as expected. At higher densities similar effects as in symmetric nuclear matter
and pure neutron matter are seen with a strong stiffening for
the models with a scalar density dependence. At a density of
$n_{b}\approx 0.11$~fm$^{-3}$ all curves of the symmetry energy cross close
to a single point and the uncertainty band is most narrow.
A similar feature was seen already for the EoS of pure neutron
matter. This particular density marks the density of highest
sensitivity in nuclei
that is below the saturation density of nuclear matter.
This crossing of EoS was already
observed very early on for Skyrme models \cite{Brown:2000pd}.

Taking the $\delta$ meson into account in the parametrisations
has almost no effect
on the EoS of symmetric or pure nucleon matter and the density dependence of
the symmetry energy.

\subsection{Dirac effective masses and scalar densities}

\begin{figure}
\resizebox{0.48\textwidth}{!}{%
  \includegraphics{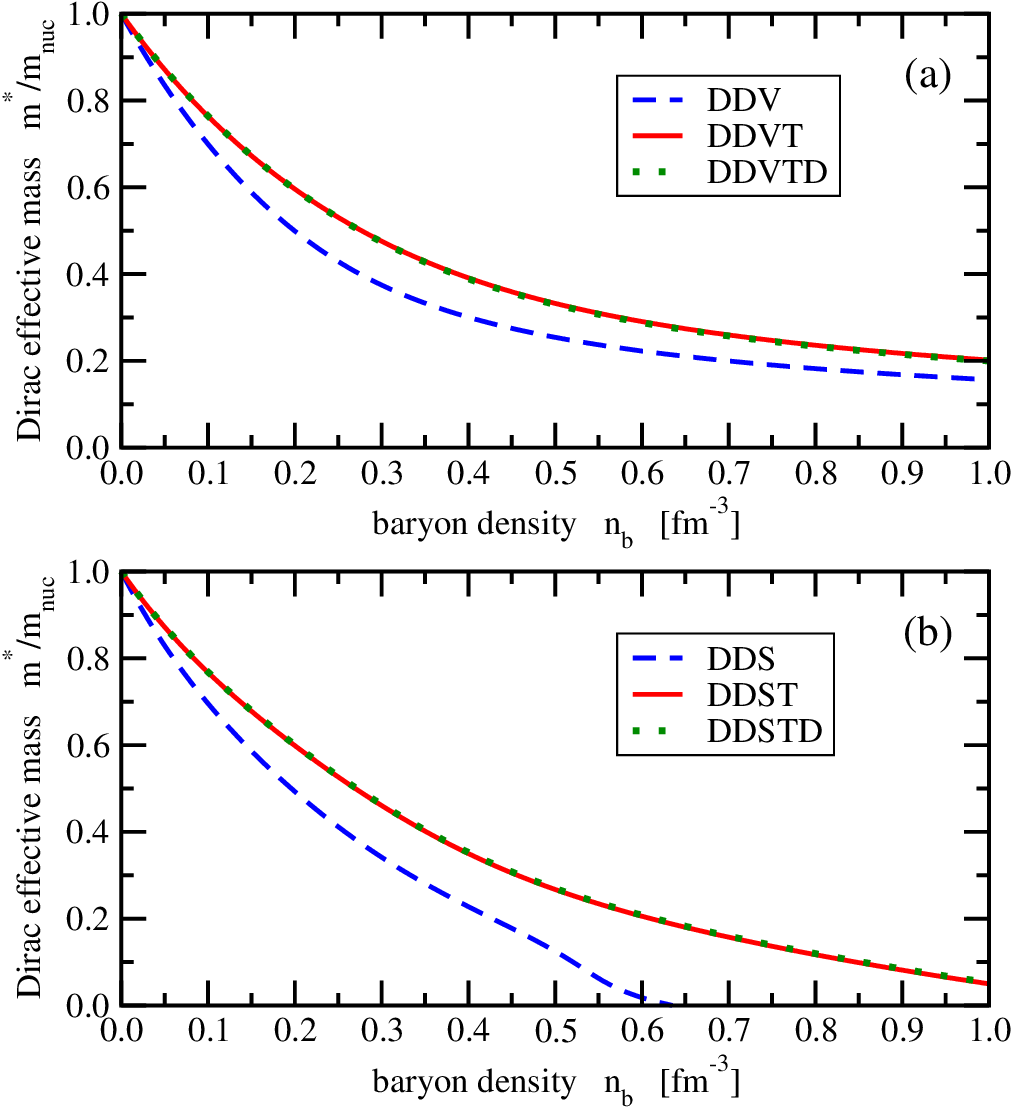}
}
\caption{Dirac effective mass in symmetric nuclear matter
  as a function of the baryon density with
  coupling functions depending on the vector density (a) and on
  the scalar density (b).}
\label{fig:07}
\end{figure}

\begin{figure}
\resizebox{0.48\textwidth}{!}{%
  \includegraphics{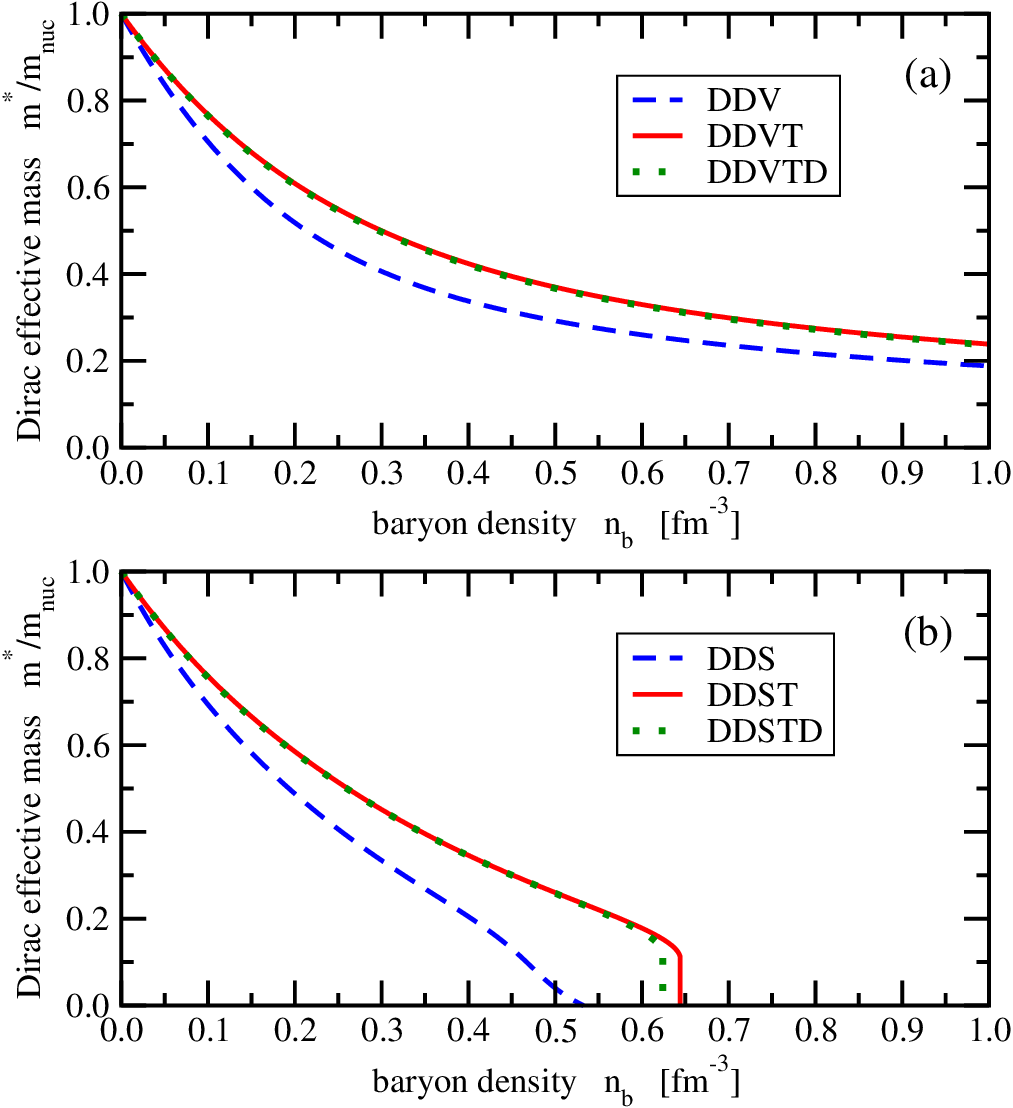}
}
\caption{Dirac effective mass in pure neutron matter
  as a function of the baryon density with
  coupling functions depending on the vector density (a) and on
  the scalar density (b).}
\label{fig:08}
\end{figure}

The Dirac effective masses $m^{\ast}_{\rm nuc}$ of the nucleons
in symmetric nuclear matter and pure neutron
matter are depicted in figures \ref{fig:07} and \ref{fig:08},
respectively. They decrease
with increasing baryon density due to the increase of the scalar
potential.
The effective masses in pure neutron matter are generally lower
than those in symmetric nuclear matter.
The models DDVT, DDVTD, DDST, and DDSTD show a larger
effective mass than the DDV and DDS models, respectively, throughout
the covered range of baryon densities in the figures. The scalar potential
$S$ has to be less strong to obtain the spin-orbit splitting in nuclei
because the tensor couplings give an additional contribution.

The models with scalar density dependence of the couplings exhibit
the peculiar feature that the Dirac effective mass vanishes at some
density above saturation. At that point the scalar potential $S_{i}$
of a nucleon $i$ reaches the vacuum nucleon mass $m_{i}$. This can happen
only because the rearrangement contribution from the $\omega$ meson
coupling in (\ref{eq:SR}) increases without bounds for a negative
derivative $d\Gamma{\omega}/dn^{(s)}$ of the $\omega$ coupling function.
The effect is more pronounced for pure neutron matter than for symmetric
nuclear matter with lower collapse densities.

\begin{figure}
\resizebox{0.48\textwidth}{!}{%
  \includegraphics{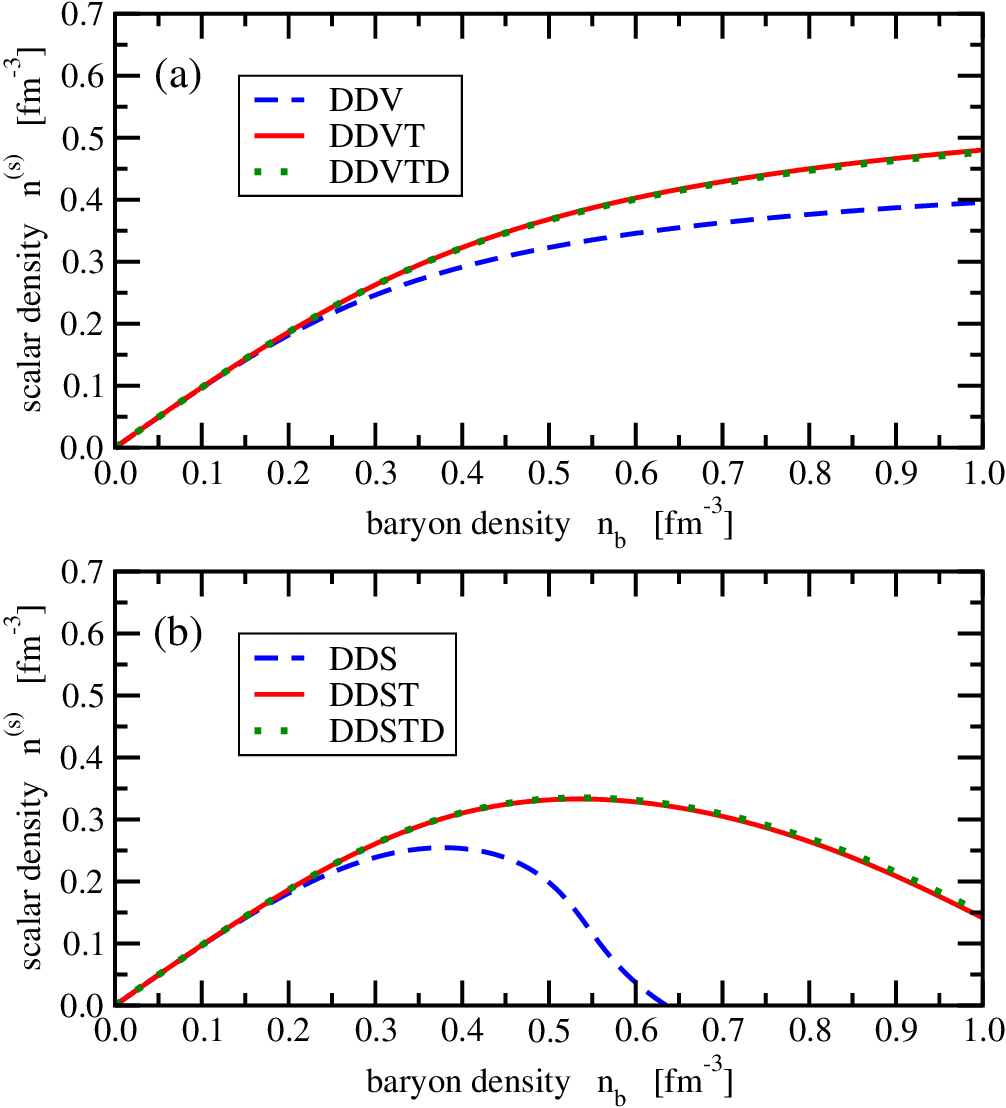}
}
\caption{Scalar density in symmetric nuclear matter
  as a function of the baryon density with
  coupling functions depending on the vector density (a) and on
  the scalar density (b).}
\label{fig:09}
\end{figure}

\begin{figure}
\resizebox{0.48\textwidth}{!}{%
  \includegraphics{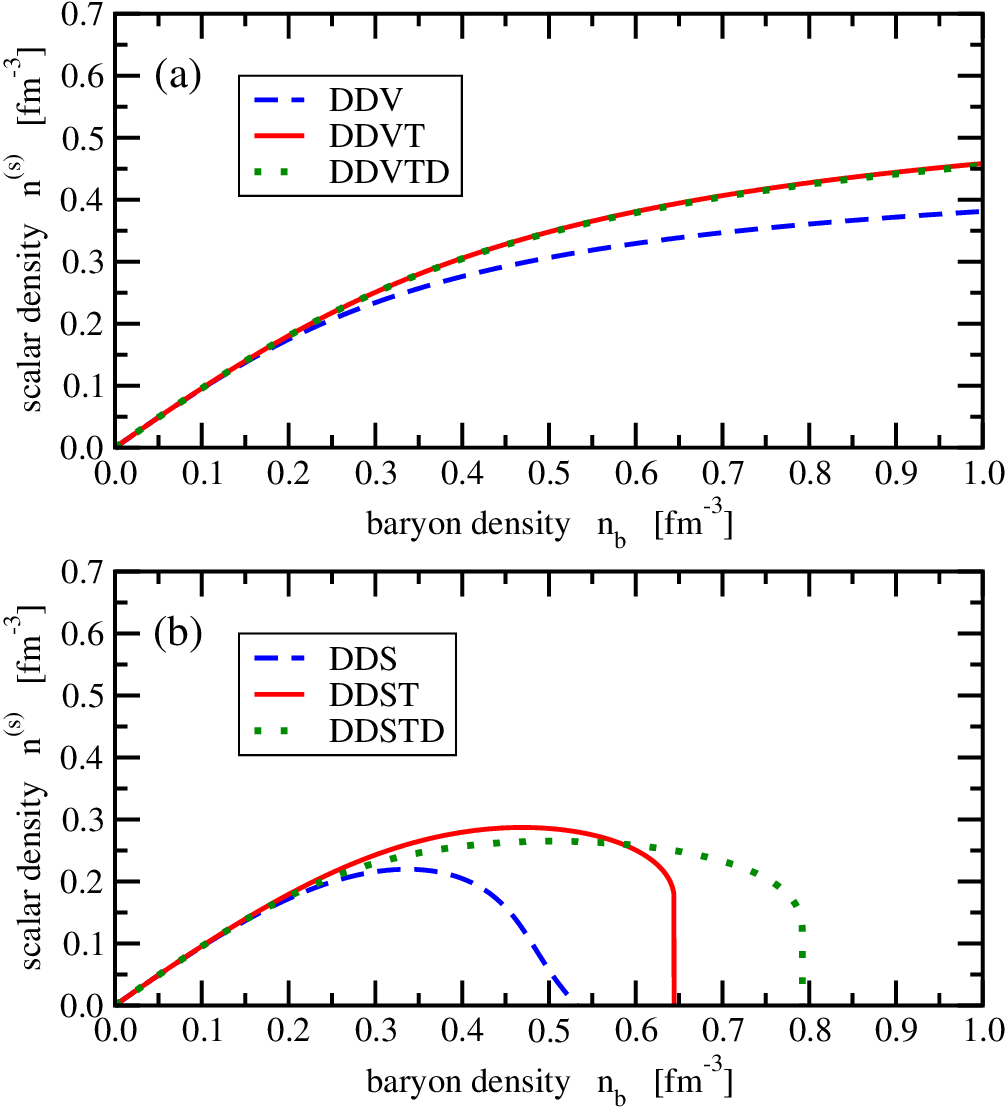}
}
\caption{Scalar density in pure neutron matter
  as a function of the baryon density with
  coupling functions depending on the vector density (a) and on
  the scalar density (b).}
\label{fig:10}
\end{figure}

The behaviour of the Dirac effective masses is correlated
with vanishing scalar densities at a certain
baryon density above saturation, as depicted in figures \ref{fig:09}
and \ref{fig:10}. In the case of models with couplings that depend
on the vector density, there is a smooth and continuous rise of the
scalar densities with decreasing slope (upper panels).
In contrast, the scalar density
in the models with scalar density dependent couplings first rise and
then decline reaching zero eventually (lower panels).

The decrease of the effective masses and scalar densities at high densities
for the models with a scalar density dependence
of the couplings modifies the kinetic
contributions (\ref{eq:pkin}) to the pressure of the system. In other models,
the decline of $m_{i}^{\ast}$ with baryon density is partly compensated by a rise
of $n_{i}^{(s)}$. In contrast, the product $m_{i}^{\ast}n_{i}^{(s)}$ in models
with couplings of scalar density dependence reduces much faster at higher
baryon densities and a stiffer EoS is expected as confirmed in figures
(\ref{fig:04}) and (\ref{fig:05}).

\section{Conclusions}
\label{sec:concl}

The construction of energy density functionals resulting from relativistic
mean-field models leaves ample room for variations. In this work two main
aspects of the form of the EDF were studied in more detail:
the dependence of the minimal nucleon-meson couplings
on the total Lorentz vector or scalar density of the system
and the effects of tensor couplings of the isovector mesons
with the nucleons. Furthermore, the effect of including
the $\delta$ meson in addition to the standard isovector $\rho$ meson
is explored.

Since relativistic EDFs are phenomenological descriptions of finite nuclei and
nuclear matter depending on a certain number of parameters,
these quantities, which
determine the effective in-medium interaction,
cannot be related directly to the free
nucleon-nucleon interaction that is well constrained by scattering data.
Instead, the model parameters have to be
found by a fit to nuclear observables. Contrary to most earlier approaches,
a self-consistent determination of the uncertainties in the $\chi^{2}$ function
was realized in this work leading to more realistic values than from heuristic
estimates.

The main results of this work can be summarized as follows. The inclusion of
tensor couplings in the relativistic EDF
reduces the strength of the isoscalar minimal nucleon-meson couplings
as compared to models without tensor couplings. For all mesons a reduction
of the couplings with increasing baryon density is found in all cases.
The description of finite nuclei improves with tensor couplings,
in particular for the binding energies and diffraction radii. The equation
of state of nuclear matter is very similar for all models
at sub-saturation densities.
The characteristic nuclear matter parameters at saturation and the
parameters of the EDFs exhibit some correlation with the choice of the
density dependence of the couplings and the inclusion of
the tensor couplings or not. Saturation densities, binding energies
at saturation, symmetry energies and slope
parameters of the models are rather well constrained. Higher-order derivatives
show much larger uncertainties.
At high densities large variations of the predictions of the EoS
are observed with a substantial
increase of the model uncertainties. The lowest uncertainties are found close
to a density of about $0.11$~fm$^{-3}$ where all EoSs cross close
to a single point. Models with a scalar density dependence are stiffer
at high densities than models
with a vector density dependence of the couplings.
This observation is related to the collapse of the total scalar density
of the system and vanishing effective mass at high baryon densities
in approaches with couplings depending on scalar densities.
The rearrangement contribution 
from the dependence of the $\omega$ coupling on
the scalar density leads to an unlimited increase of the
scalar potential in this case. Models with tensor couplings have
larger effective nucleon masses than models without.
This is possible because the tensor contributions increase the
spin-orbit splitting so that a smaller strength of the
Lorentz scalar $\sigma$ meson is needed to fit the nuclear data.

In order to avoid the problems with the scalar density dependence
of all couplings,
a change to a mixed density dependence can be considered, i.e.\ a dependence of
Lorentz vector meson couplings on the vector density and Lorentz scalar meson
couplings on the scalar density. Models of this type have been explored in
\cite{Typel:2018cap} but without tensor couplings and without a self-consistent
determination of the uncertainties of the fitted observables.
Work in this direction is in progress using an extended set of
nuclei and more observables in the fitting procedure.

Another aspect is also worthwhile to be investigated further.
In the present approach,
the incompressibility of nuclear matter was kept to a fixed value because a fit
to only ground state properties of finite nuclei does not help
to fix this quantity
very well. Properties of excited states like giant resonances,
in particular the isoscalar giant monopole resonance, 
have to be included in the fitting procedure and 
can help much better to find proper values of $K$, see, e.g.,
\cite{Shlomo:2019bgh}. The study of other types of giant resonances
that constrain the isovector part of the effective interaction will
also be useful.

In general, a better description of surface properties of finite nuclei
will be very rewarding by considering appropriately chosen
new terms in the EDF. It is known that the incompressibility
of nuclear matter is closely related to the surface thickness and
surface tension as expressed by the so-called pocket formula
\cite{Stocker:1980olw,Brack:1982yki}. More precise experimental
data on the neutron skin thickness would help to determine
the isovector parts of the EDF more precisely.

\section{Acknowledgements}
Discussions with Daniel R.\ Phillips, Shalom Shlomo, and Dario Vretenar
are acknowledged. This work was initiated during a two-month visit
of DAT at the Institut f\"{u}r Kernphysik in Darmstadt which
was supported by an Erasmus+ traineeship.

\section*{Appendix A}

The rearrangement contributions (\ref{eq:SR}) and (\ref{eq:VR}) to the
vector and scalar potentials (\ref{eq:S}) and (\ref{eq:V}), respectively,
arise from the density dependence of the couplings $\tilde{\Gamma}_{j}$ in the
covariant derivative (\ref{eq:D}) and the effective mass operator (\ref{eq:M})
when the field equation of the nucleons is deduced with the help of the
Euler-Langrange equation
\begin{equation}
  \label{eq:EL}
  0 = \frac{\partial \mathcal{L}}{\partial \overline{\Psi}_{ik}}
  - \partial_{\nu}\frac{\partial \mathcal{L}}{\partial
    (\partial_{\nu}\overline{\Psi}_{ik})} 
\end{equation}
with the Lagrangian density (\ref{eq:L}). Here the indices $i$ and $k$ are used
with $i$ distinguishing between protons and neutrons and $k$ to identify a
particular single-particle state. Hence the nucleonic contribution
to $\mathcal{L}$ reads
\begin{eqnarray}
  \lefteqn{\mathcal{L}_{\rm nucleon}}
  \\ \nonumber & = &
  \sum_{i=p,n} \sum_{k} w_{ik} \overline{\Psi}_{ik}
 \left( \gamma_{\mu}iD_{ik}^{\mu} - \sigma_{\mu\nu}T_{ik}^{\mu\nu}- M_{ik}^{\ast} \right) \Psi_{ik}
\end{eqnarray}
if the occupation factors $w_{ik}=0$ or $1$ are introduced similar as in the definitions
(\ref{eq:curr}) and (\ref{eq:rho_s}) of the current and scalar density.
Since $\mathcal{L}$ does not depend
on derivatives of $\overline{\Psi}_{ik}$ only the first term in (\ref{eq:EL})
remains. The result
\begin{eqnarray}
   \label{eq:EL2}
  \lefteqn{0  =  w_{ik}
  \left( \gamma_{\mu}iD_{ik}^{\mu} - \sigma_{\mu\nu}T_{ik}^{\mu\nu}
  - M_{ik}^{\ast} \right) \Psi_{ik}}
  \\ \nonumber & &
  - \sum_{i^{\prime}=p,n} \sum_{k^{\prime}} w_{i^{\prime}k^{\prime}}
    \overline{\Psi}_{i^{\prime}k^{\prime}} \gamma_{\mu}
  \left( \frac{\partial \tilde{\Gamma}_{\omega}}{\partial \overline{\Psi}_{ik}}
  \omega^{\mu} +\frac{\partial \tilde{\Gamma}_{\varrho}}{\partial \overline{\Psi}_{ik}}
  \vec{\varrho}^{\mu} \cdot \vec{\tau}
  \right) \Psi_{i^{\prime}k^{\prime}}
  \\ \nonumber & &
  + \sum_{i^{\prime}=p,n} \sum_{k^{\prime}}w_{i^{\prime}k^{\prime}}
  \overline{\Psi}_{i^{\prime}k^{\prime}} 
  \left( \frac{\partial \tilde{\Gamma}_{\sigma}}{\partial \overline{\Psi}_{ik}}
  \sigma + \frac{\partial \tilde{\Gamma}_{\delta}}{\partial \overline{\Psi}_{ik}}
  \vec{\delta} \cdot \vec{\tau}  \right) \Psi_{i^{\prime}k^{\prime}}
\end{eqnarray}
contains in the first line the usual form of the Dirac equation for nucleon $k$.
The dependence of the couplings $\tilde{\Gamma}_{j}$
on the nucleon fields leads to the contributions in the second and third line.
Assuming a dependence on the vector density $\rho_{v}$ or the scalar density
$\rho_{s}$, the expression
\begin{eqnarray}
  \frac{\partial \tilde{\Gamma}_{j}}{\partial \overline{\Psi}_{ik}}
  & = & \frac{\partial \tilde{\Gamma}_{j}}{\partial \rho_{v}}
  \frac{\partial \rho_{v}}{\partial \overline{\Psi}_{ik}}
  + \frac{\partial \tilde{\Gamma}_{j}}{\partial \rho_{s}}
  \frac{\partial \rho_{s}}{\partial \overline{\Psi}_{ik}}
  \\ \nonumber & = &
  \frac{\partial \tilde{\Gamma}_{j}}{\partial \rho_{v}}
  \frac{j^{\mu}}{\rho_{v}} \gamma_{\mu} w_{ik} \Psi_{ik}
  + \frac{\partial \tilde{\Gamma}_{j}}{\partial \rho_{s}}
  w_{ik} \Psi_{ik}
\end{eqnarray}
is found
with equations (\ref{eq:curr}) and (\ref{eq:rho_s}). The equation (\ref{eq:EL2})
becomes
\begin{eqnarray}
   \label{eq:EL3}
   0 & = &
   \gamma_{\mu}\left(  i \partial^{\mu} - \tilde{\Gamma}_{\omega} \omega^{\mu}
  - \tilde{\Gamma}_{\rho} \vec{\rho}^{\mu} \cdot \vec{\tau}
  - \Gamma_{\gamma}  A^{\mu} \frac{1+\tau_{3}}{2}
 \right) \Psi_{ik}
  \\ \nonumber & &
   -  \left( m_{i} - \tilde{\Gamma}_{\sigma} \sigma 
 - \tilde{\Gamma}_{\delta} \vec{\delta} \cdot \vec{\tau} \right) \Psi_{ik}
  \\ \nonumber & &
  -  \left(
 \frac{\partial \tilde{\Gamma}_{\omega}}{\partial \rho_{v}}
  \frac{j^{\mu}}{\rho_{v}} \gamma_{\mu}
  + \frac{\partial \tilde{\Gamma}_{\omega}}{\partial \rho_{s}}  \right)
  \Psi_{ik} \omega^{\nu}  \sum_{i^{\prime}=p,n} \sum_{k^{\prime}}
  \overline{\Psi}_{i^{\prime}k^{\prime}} \gamma_{\nu} \Psi_{i^{\prime}k^{\prime}}
  \\ \nonumber & & 
  -   \left(
   \frac{\partial \tilde{\Gamma}_{\rho}}{\partial \rho_{v}}
  \frac{j^{\mu}}{\rho_{v}} \gamma_{\mu}
  + \frac{\partial \tilde{\Gamma}_{\rho}}{\partial \rho_{s}} \right) 
  \Psi_{ik} \vec{\varrho}^{\nu}
  \cdot \sum_{i^{\prime}=p,n} \sum_{k^{\prime}}\overline{\Psi}_{i^{\prime}k^{\prime}}
  \gamma_{\nu} \vec{\tau} \Psi_{i^{\prime}k^{\prime}}
  \\ \nonumber & &
  +   \left(  \frac{\partial \tilde{\Gamma}_{\sigma}}{\partial \rho_{v}}
  \frac{j^{\mu}}{\rho_{v}} \gamma_{\mu}
  + \frac{\partial \tilde{\Gamma}_{\sigma}}{\partial \rho_{s}}  \right)
  \Psi_{ik}   \sigma  \sum_{i^{\prime}=p,n}  \sum_{k^{\prime}} \overline{\Psi}_{i^{\prime}k^{\prime}}
  \Psi_{i^{\prime}k^{\prime}}
    \\ \nonumber & &
  +   \left(  \frac{\partial \tilde{\Gamma}_{\delta}}{\partial \rho_{v}}
  \frac{j^{\mu}}{\rho_{v}} \gamma_{\mu}
  + \frac{\partial \tilde{\Gamma}_{\delta}}{\partial \rho_{s}} \right)
  \Psi_{ik}   \vec{\delta} \cdot \sum_{i^{\prime}=p,n}  \sum_{k^{\prime}}
  \overline{\Psi}_{i^{\prime}k^{\prime}}  \vec{\tau} \Psi_{i^{\prime}k^{\prime}}
  \\ \nonumber & &
  - \sigma_{\mu\nu}T_{i}^{\mu\nu} \Psi_{ik}
\end{eqnarray}
where the source currents and densities
\begin{eqnarray}
  \label{eq:jomega}
  j_{\omega\mu} & = & \sum_{i^{\prime}=p,n}  \sum_{k^{\prime}} w_{i^{\prime}k^{\prime}}
  \overline{\Psi}_{i^{\prime}k^{\prime}} \gamma_{\mu} \Psi_{i^{\prime}k^{\prime}}
  \\
  \label{eq:jrho}
  \vec{j}_{\rho\mu} & = & \sum_{i^{\prime}=p,n}  \sum_{k^{\prime}} w_{i^{\prime}k^{\prime}}
  \overline{\Psi}_{i^{\prime}k^{\prime}} \gamma_{\mu} \vec{\tau} \Psi_{i^{\prime}k^{\prime}}
  \\
  \label{eq:rhosigma}
  \varrho_{\sigma} & = & \sum_{i^{\prime}=p,n}  \sum_{k^{\prime}}  w_{i^{\prime}k^{\prime}}
  \overline{\Psi}_{i^{\prime}k^{\prime}} \Psi_{i^{\prime}k^{\prime}}
  \\
  \label{eq:rhodelta}
  \vec{\varrho}_{\delta} & = & \sum_{i^{\prime}=p,n}  \sum_{k^{\prime}} w_{i^{\prime}k^{\prime}}
  \overline{\Psi}_{i^{\prime}k^{\prime}}  \vec{\tau} \Psi_{i^{\prime}k^{\prime}}
\end{eqnarray}
are easily identified. Reordering the contributions gives
\begin{eqnarray}
   \label{eq:EL4}
  0 & = & \left[
   \gamma_{\mu}  \left( i \partial^{\mu}  -  \Sigma^{\mu} \right)
    - \sigma_{\mu\nu}T_{i}^{\mu\nu} 
    - \left( m_{i}  - \Sigma \right) \right] \Psi_{ik}
\end{eqnarray}
with the explicit forms of the scalar self-energy
\begin{eqnarray}
  \label{eq:sigmas}
  \Sigma & = &    \tilde{\Gamma}_{\sigma} \sigma
  +\frac{\partial \tilde{\Gamma}_{\sigma}}{\partial \rho_{s}}
   \sigma \varrho_{\sigma}
  + \tilde{\Gamma}_{\delta} \vec{\delta} \cdot \vec{\tau}
  +\frac{\partial \tilde{\Gamma}_{\delta}}{\partial \rho_{s}}
  \vec{\delta} \cdot \vec{\varrho}_{\delta}
  \\ \nonumber & & 
  -  \frac{\partial \tilde{\Gamma}_{\omega}}{\partial \rho_{s}}
  \omega^{\nu} j_{\omega\nu}
  -  \frac{\partial \tilde{\Gamma}_{\rho}}{\partial \rho_{s}}
  \vec{\varrho}^{\nu}  \cdot  \vec{j}_{\rho\nu} 
\end{eqnarray}
and the vector self-energy
\begin{eqnarray}
  \label{eq:sigmav}
  \Sigma^{\mu} & = &  \tilde{\Gamma}_{\omega} \omega^{\mu}
  + \frac{\partial \tilde{\Gamma}_{\omega}}{\partial \rho_{v}}
  \frac{j^{\mu}}{\rho_{v}}\omega^{\nu} j_{\omega\nu}
  + \tilde{\Gamma}_{\rho} \vec{\rho}^{\mu} \cdot \vec{\tau}
  + \frac{\partial \tilde{\Gamma}_{\rho}}{\partial \rho_{v}}
  \frac{j^{\mu}}{\rho_{v}}
  \\ \nonumber & & 
  - \frac{\partial \tilde{\Gamma}_{\sigma}}{\partial \rho_{v}}
  \frac{j^{\mu}}{\rho_{v}}\sigma \varrho_{\sigma}
  -  \frac{\partial \tilde{\Gamma}_{\delta}}{\partial \rho_{v}}
  \frac{j^{\mu}}{\rho_{v}}  \vec{\delta} \cdot \vec{\varrho}_{\delta}
   + \Gamma_{\gamma}  A^{\mu} \frac{1+\tau_{3}}{2} \: .
\end{eqnarray}
In the mean-field approximation and under the symmetries of the application
to spherical nuclei, the currents and densities (\ref{eq:jomega}),
(\ref{eq:jrho}), (\ref{eq:rhosigma}), and (\ref{eq:rhodelta}) reduce
to the forms (\ref{eq:nv}) and (\ref{eq:ns}) with the appropriate
coupling factors $g_{ij}$. Similarly, the self-energies (\ref{eq:sigmas}) and
(\ref{eq:sigmav}) become simple potentials (\ref{eq:S}) and (\ref{eq:V})
where only the $\mu=0$ component of $\Sigma^{\mu}$ and $j^{0}=\rho_{v}$
remain. The rearrangement
contributions (\ref{eq:SR}) and (\ref{eq:VR}) are easily recognized when
the functionals $\tilde{\Gamma}_{j}$ of the fields
are replaced by functions $\Gamma_{j}$ of the densities.

\section*{Appendix B}
\label{app:b}

The energy density functional (\ref{eq:eps}) contains in total $19$
parameters: the masses of nucleons and mesons (six), the reference density (one)
and the nucleon-meson couplings at this density (four), the parameters
describing the density dependence of the couplings (six), and the
tensor couplings (two). However, really used
are only $14$ because the masses of the nucleons and mesons, except
for the $\sigma$ meson, are kept constant. Thus there are eight parameters
for the isoscalar part of the effective interaction and six for the
isovector part.

In the actual fitting procedure six of the eight isoscalar parameters are
replaced by four characteristic parameters of symmetric nuclear matter
and two auxiliary quantities. The mass of the sigma meson $m_{\sigma}$
and the $\omega$ tensor coupling $\Gamma_{T\omega}$ are used directly
as independent quantities. The four nuclear matter parameters are
the saturation density $n_{\rm sat}$, the Dirac effective mass $m^{\ast}_{\rm sat}$
at saturation, the binding energy per nucleon at saturation $B_{\rm sat}$
and the incompressibility $K$. The two auxiliary quantities are
the derivative $f_{\omega}^{\prime}(1)$
and the ratio
$r = f_{\omega}^{\prime\prime}(1)/f_{\omega}^{\prime}(1)$
of derivatives. The transformation proceeds in two steps. First, the nuclear
matter parameters are converted to the coupling factors
$C_{j}=\Gamma_{j}^{2}/m_{j}^{2}$ for $j=\omega,\sigma$, their first
derivatives $D_{j}^{(v)}$ or $D_{j}^{(s)}$ and second derivatives
$E_{j}^{(v)}$ or $E_{j}^{(s)}$ as defined in subsection \ref{sec:edf}.
Then these quantities are used
to calculated the parameters $a_{j}$, $b_{j}$, $c_{j}$, and $d_{j}$
of the rational function (\ref{eq:fso}).

In the parameter conversion, the average
nucleon mass $m_{\rm nuc}=(m_{p}+m_{n})/2$ is used and the real nuclear matter
parameters are calculated numerically at the end of the fitting procedure
with the correct nucleon masses to be independent of this averaging.

The saturation density $n_{\rm sat}$ defines the Fermi momentum
\begin{equation}
  p_{\rm sat}^{\ast} = \left( \frac{3\pi^{2}}{2} n_{\rm sat}\right)^{\frac{1}{3}}
\end{equation}
in symmetric nuclear matter. With the Dirac effective mass $m_{\rm sat}^{\ast}$ the
effective chemical potential
\begin{equation}
  \mu_{\rm sat}^{\ast}=\sqrt{(p_{\rm sat}^{\ast})^{2}+(m_{\rm sat}^{\ast})^{2}}
\end{equation}
and the corresponding scalar density
\begin{equation}
  n_{\rm sat}^{(s)} =
  \frac{m_{\rm sat}^{\ast}}{\pi^{2}} \left[ p_{\rm sat}^{\ast}\mu_{\rm sat}^{\ast}
    - \left( m_{\rm sat}^{\ast} \right)^{2}
    \ln \frac{p_{\rm sat}^{\ast}+\mu_{\rm sat}^{\ast}}{m_{\rm sat}^{\ast}} \right]
\end{equation}
at saturation are given. The saturation density $n_{\rm sat}$ or the
scalar density at saturation $n_{\rm sat}^{(s)}$ are used as the
reference density $n_{\rm ref}$ in the coupling functions if
a vector or scalar density dependence is selected, respectively.
Then the scalar and vector potential at saturation
are found as
\begin{equation}
  S_{\rm sat} = m_{\rm nuc} - m_{\rm sat}^{\ast}
\end{equation}
and
\begin{equation}
  V_{\rm sat} = m_{\rm nuc} - B_{\rm sat} - \mu_{\rm sat}^{\ast}
\end{equation}
with the binding energy per nucleon $B_{\rm sat}$. Furthermore the
factor (\ref{eq:facf}) at saturation
\begin{equation}
  f_{\rm sat} = 3 \left( \frac{n_{\rm sat}^{(s)}}{m_{\rm sat}^{\ast}}
  - \frac{n_{\rm sat}}{\mu_{\rm sat}^ {\ast}}\right)
\end{equation}
is defined for later convenience. For the next step
one has to distinguish the type of density dependence.

In case of a vector density dependence with $n_{\rm ref} = n_{\rm sat}^{(v)}$
there is no rearrangement
contribution to the scalar potential, $S_{\rm sat}^{(R)}=0$,
and the $\sigma$ coupling
\begin{equation}
  C_{\sigma}(n_{\rm ref}) = \frac{\Gamma_{\sigma}^{2}(n_{\rm ref})}{m_{\sigma}^{2}}
  = \frac{S_{\rm sat}}{n_{\rm sat}^{(s)}}
\end{equation}
and thus $\Gamma_{\sigma}(n_{\rm ref})$ for given $m_{\sigma}$
are immediately obtained. The vector coupling is found with (\ref{eq:eps})
to be
\begin{eqnarray}
  C_{\omega}(n_{\rm ref}) & = & \frac{\Gamma_{\omega}^{2}(n_{\rm ref})}{m_{\omega}^{2}}
  \\ \nonumber   & = & \frac{2}{n_{\rm sat}^{2}}
  \left[ \left(m_{\rm nuc}-B_{\rm sat}\right)n_{\rm sat}
    - \frac{1}{2} C_{\sigma}\left( n_{\rm sat}^{(s)}\right)^{2}
    \right.  \\ \nonumber & & \left.
    - \frac{1}{4} \left( 3 \mu_{\rm sat}^{\ast}n_{\rm sat}
    + m_{\rm sat}^{\ast}n_{\rm sat}^{\ast}\right)
    \right]
\end{eqnarray}
in symmetric nuclear matter. Then the rearrangement contribution to
$V_{\rm sat}$ is
\begin{equation}
  V_{\rm sat}^{(R)} = V_{\rm sat} - C_{\omega}\left(n_{\rm sat}^{(v)}\right)^{2}
\end{equation}
and the derivatives
\begin{equation}
  D_{\omega}^{(v)}(n_{\rm ref}) = 2 f_{\omega}^{\prime}(1)
  \frac{C_{\omega}}{n_{\rm sat}^{(v)}}
\end{equation}
and
\begin{equation}
  D_{\sigma}^{(v)}(n_{\rm ref}) =
  \frac{D_{\omega}^{(v)}\left(n_{\rm ref}\right)^{2}
  -2 V_{\rm sat}^{(R)}}{\left(n_{\rm sat}^{(s)}\right)^{2}}
\end{equation}
are determined with the auxiliary quantity $f_{\omega}^{\prime}(1)$.
Defining
\begin{equation}
  E_{\omega}^{(v)}(n_{\rm ref}) = \frac{2C_{\omega}}{(n_{\rm sat}^{(v)})^{2}}
  f_{\omega}^{\prime}(1)
  \left[  f_{\omega}^{\prime}(1) + r \right]
\end{equation}
and
\begin{equation}
  g_{\rm sat}^{(v)} = \frac{\frac{m_{\rm sat}^{\ast}}{\mu_{\rm sat}^{\ast}}
  - f_{\rm sat} D_{\sigma}^{(v)}n_{\rm sat}^{(s)}}{1+f_{\rm sat}C_{\sigma}}
\end{equation}
the second derivative
\begin{eqnarray}
  \lefteqn{E_{\sigma}^{(v)}(n_{\rm ref}) =}
  \\ \nonumber & &
  \left[ 3 \left( \mu_{\rm sat}^{\ast}-m_{\rm sat}^{\ast}g_{\rm sat}^{(v)}\right) -K
    \right.  \\ \nonumber & &
    + 9 C_{\omega}n_{\rm sat}^{(v)}
    +18 D_{\omega}^{(v)}(n_{\rm sat}^{(v)})^{2}
    + \frac{9}{2} E_{\omega}^{(v)}(n_{\rm sat}^{(v)})^{3}
    \\ \nonumber & & \left.
    - 9 D_{\sigma}^{(v)} \left(n_{\rm sat}^{(s)}\right)^{2}
    - 9 \left( C_{\sigma} + D_{\sigma}^{(v)}n_{\rm sat}^{(v)}\right)
    n_{\rm sat}^{(s)}g_{\rm sat}^{(v)}\right]
  \\ \nonumber & &
   \left[ \frac{9}{2} n_{\rm sat}^{(v)} \left(n_{\rm sat}^{(s)}\right)^{2}\right]^{-1}
\end{eqnarray}
is found from the incompressibility $K$
using (\ref{eq:dpdnv}) and (\ref{eq:dnsdnv}).

The calculation proceeds slightly differently in case of couplings that depend
of the scalar density. 
It is more complicated because both expressions
(\ref{eq:dpdnv}) and (\ref{eq:dnsdnv}) depend explicitly on the
second derivatives of the couplings.
Since the rearrangement contribution $V_{\rm sat}^{(R)}$
is zero, the $\omega$ coupling
\begin{equation}
  C_{\omega}(n_{\rm ref}) = \frac{\Gamma_{\omega}^{2}(n_{\rm ref})}{m_{\omega}^{2}}
  = \frac{V_{\rm sat}}{n_{\rm sat}^{(v)}}
\end{equation}
is directly found as well as the derivatives
\begin{equation}
  D_{\omega}^{(v)}(n_{\rm ref}) = 2 f_{\omega}^{\prime}(1)
  \frac{C_{\omega}}{n_{\rm ref}}
\end{equation}
and
\begin{equation}
  E_{\omega}^{(v)}(n_{\rm ref}) = \frac{2C_{\omega}}{n_{\rm ref}^{2}}
  f_{\omega}^{\prime}(1)
  \left[  f_{\omega}^{\prime}(1) + r \right]
\end{equation}
where the reference density $n_{\rm ref}$ is now identical to the
scalar density $n_{\rm sat}^{(s)}$. The rearrangement contribution
to the scalar potential is
\begin{eqnarray}
  S_{\rm sat}^{(R)} & = & \frac{1}{n_{\rm sat}^{(s)}}
  \left( \frac{1}{2} \mu_{\rm sat}^{\ast}n_{\rm sat}^{(v)}
  -\frac{1}{2} m_{\rm sat}^{\ast}n_{\rm sat}^{(s)}
  \right. \\ \nonumber & & \left.
  +  V_{\rm sat} n_{\rm sat}^{(v)}
  -  S_{\rm sat} n_{\rm sat}^{(s)}\right)
\end{eqnarray}
because the pressure (\ref{eq:pres}) is zero at the saturation density.
If one defines the auxiliary quantity
\begin{equation}
  X_{\rm sat} = \frac{Y_{\rm sat}}{Z_{\rm sat}}
\end{equation}
as the ratio of
\begin{eqnarray}
  Y_{\rm sat}   & = &\left(  \frac{m_{\rm sat}^{\ast}}{\mu_{\rm sat}^{\ast}} 
  +f_{\rm sat} D_{\omega}^{(s)} n_{\rm sat}^{(v)} \right)
    \\ \nonumber & & 
  \left( C_{\sigma}  n_{\rm sat}^{(s)}
  +  2 D_{\sigma}^{(s)} (n_{\rm sat}^{(s)})^{2}
  -  D_{\omega}^{(s)} (n_{\rm sat}^{(v)})^{2}
  + \frac{m_{\rm sat}^{\ast}}{3}  \right)
  \\ \nonumber & &
  -   \left[ 1  +   f_{\rm sat}\left(  C_{\sigma}
    + 2 D_{\sigma}^{(s)} n_{\rm sat}^{(s)}  \right) \right]
  \\ \nonumber & &
  \left[\frac{\mu_{\rm sat}^{\ast}}{3} -  \frac{K}{9}
    +  \left( C_{\omega} +  D_{\omega}^{(s)}n_{\rm sat}^{(s)} \right)
    n_{\rm sat}^{(v)} \right]
\end{eqnarray}
and
\begin{equation}
  Z_{\rm sat}  =  f_{\rm sat}\left(\frac{\mu_{\rm sat}^{\ast}}{3} -  \frac{K}{9}
    +  C_{\omega}  n_{\rm sat}^{(v)} \right)
  -   n_{\rm sat}^{(s)}  \frac{m_{\rm sat}^{\ast}}{\mu_{\rm sat}^{\ast}} 
\end{equation}
then the second derivative $E_{\sigma}^{(s)}(n_{\rm ref})$
is easily obtained from
\begin{equation}
  E_{\sigma}^{(s)} =
  \frac{1}{(n_{\rm sat}^{(s)})^{2}}
  \left[2 X_{\rm sat}  + E_{\omega}^{(s)}  (n_{\rm sat}^{(v)})^{2} \right] \: .
\end{equation}

In the next step of the parameter conversion, the couplings $C_{\sigma}$
and their derivatives are used to define the function derivatives
\begin{equation}
  f_{\sigma}^{\prime}(1) = \frac{D_{\sigma}^{(x)}n_{\rm sat}^{(x)}}{2C_{\sigma}}
\end{equation}
and
\begin{equation}
  f_{\sigma}^{\prime\prime}(1) = \frac{E_{\sigma}^{(x)}(n_{\rm sat}^{(x)})^{2}}{2C_{\sigma}}
  - \left[ f_{\sigma}^{\prime}(1)\right]^{2}
\end{equation}
where $x=v$ or $s$ depending on the type of density dependence.
The corresponding
values $f_{\omega}^{\prime}(1)$ and $f_{\omega}^{\prime\prime}(1)$ are already known
because they were used as original parameters in the fit.

Finally the quantities $f_{j}^{\prime}(1)$ and $f_{j}^{\prime\prime}(1)$
for $j=\omega$ and $\sigma$ are used to determine the actual parameters of
the rational function (\ref{eq:fso}). For this purpose, the first and second
derivatives of the function (\ref{eq:fso}) are needed. They are given
by
\begin{equation}
\label{eq:fpso}
f_{j}^{\prime}(x) = 2 a_{j} (b_{j}-c_{j})
\frac{(x+d_{j})}{\left[1+c_{j}(x+d_{j})^{2}\right]^{2}}
\end{equation}
and
\begin{equation}
\label{eq:fppso}
f_{j}^{\prime\prime}(x) = 2 a_{j} (b_{j}-c_{j})
\frac{1-3c_{j}(x+d_{j})^{2}}{\left[1+c_{j}(x+d_{j})^{2}\right]^{3}}
\end{equation}
where the condition $f_{j}^{\prime\prime}(0)=0$ leads to the
condition $c_{j}=1/(3d_{j}^{2})$. With the known ratio
\begin{eqnarray}
  R_{21}(d_{j})  = \frac{f_{j}^{\prime\prime}(1)}{f_{j}^{\prime}(1)}
  & = & \frac{1-3c_{j}(1+d_{j})^{2}}{\left[1+c_{j}(1+d_{j})^{2}\right](1+d_{j})}
  \\ \nonumber & = &
  \frac{-1-2d_{j}}{\left[d_{j}^{2}+\frac{1}{3}(1+d_{j})^{2}\right](1+d_{j})}
\end{eqnarray}
an equation to determine $d_{j}$ is obtained. The function $R_{21}(d_{j})$
is negative for $d_{j}>-1/2$. A unique solution with positive $d_{j}$
is possible for $-3 \leq R_{21} < 0$. Only this branch is considered in
the present parameter fits. With the known $d_{j}$ and then $c_{j}$,
the parameter $b_{j}$ is found from the ratio
\begin{eqnarray}
  R_{10}(d_{j}) & = & \frac{f_{j}^{\prime}(1)}{f_{j}(1)}
  \\ \nonumber   & = & 
  \frac{2(b_{j}-c_{j})(1+d_{j})}{[1+c_{j}(1+d_{j})^{2}][1+b_{j}(1+d_{j})^{2}]}  
\end{eqnarray}
as
\begin{equation}
  b_{j} = \frac{R_{10}+c_{j}S}{S-R_{10}(1+d_{j})^{2}}
\end{equation}
with the auxiliary quantity
\begin{equation}
  S = \frac{2(1+d_{j})}{1+c_{j}(1+d_{j})^{2}} \: .
\end{equation}
Finally, the parameter $a_{j}$ is found from the normalisation condition
$f_{j}(1)=1$ as
\begin{equation}
  a_{j} = \frac{1+c_{j}(1+d_{j})^{2}}{1+b_{j}(1+d_{j})^{2}} \: .
\end{equation}

\clearpage
\newpage

 \bibliographystyle{unsrt}
 \bibliography{paper}

\clearpage
\newpage

\begin{table*}
  \caption{Mass of the $\sigma$ meson and coupling strengths
    at the reference density for the different parametrisations.}
\label{tab:01}
\begin{center}
\begin{tabular}{ccccccccc}
\hline\noalign{\smallskip}
parametrisation & number of & mass of
 & $\omega$ meson & $\sigma$ meson & $\rho$ meson & $\delta$ meson
 & $\omega$ tensor & $\rho$ tensor \\
& parameters & $\sigma$ meson
 & coupling & coupling & coupling & coupling
 & coupling & coupling \\
& & $m_{\sigma}$
 & $\Gamma_{\omega}^{(0)}$ & $\Gamma_{\sigma}^{(0)}$
 & $\Gamma_{\rho}^{(0)}$ & $\Gamma_{\delta}^{(0)}$
 & $\Gamma_{T\omega}$ & $\Gamma_{T\rho}$\\
& $N_{\rm par}$ & [MeV]
 & & & & & & \\
\noalign{\smallskip}\hline\noalign{\smallskip}
 DDV   &  8 & 537.600098 & 12.770450 & 10.136960 & 3.9241650 & 0.0000000
 & 0.000000 & 0.0000000 \\
 DDVT  & 10 & 502.598602 & 10.987106 & 8.382863 & 3.8485560 & 0.0000000
 & 3.681512 & 12.608333 \\
 DDVTD & 11 & 502.619843 & 10.980433 & 8.379269 & 4.0301900 & 0.8487420
 & 3.671994 & 12.382725 \\
 DDS   &  8 & 539.257996 & 12.925653 & 10.272845 & 3.9619391 & 0.0000000
 & 0.000000 & 0.0000000 \\
 DDST  & 10 & 506.451447 & 11.170835 & 8.556703 & 3.8558209 & 0.0000000
 & 3.869954 & 12.601035 \\
 DDSTD & 11 & 505.835999 & 11.138900 & 8.526903 & 4.0612888 & 1.7924930
 & 3.938227 & 12.087422 \\
\noalign{\smallskip}\hline
\end{tabular}
\end{center}
\end{table*}

\begin{table*}
  \caption{Parameters for the density dependence of couplings. The coefficients $d_{j}$ for $j=\omega, \sigma$
    are given by $d_{j} = 1/\sqrt{3c_{j}}$, except for
    $d_{\omega}= 3758.39866319$ in case of DDSTD.
  }
\label{tab:02}
\begin{center}
\begin{tabular}{cccccccc}
\hline\noalign{\smallskip}
parametrisation & $n_{\rm ref}^{(v)}$ & $n_{\rm ref}^{(s)}$ & $b_{\omega}$ & $c_{\omega}$
& $b_{\sigma}$ & $c_{\sigma}$ & $a_{\rho}=a_{\delta}$ \\
\noalign{\smallskip}\hline\noalign{\smallskip}
 DDV   & 0.151117 & 0.14218170 & 0.03911422 & 0.07239939 & 0.21286844 & 0.30798197 & 0.35265899 \\
 DDVT  & 0.153623 & 0.14636172 & 0.04459850 & 0.06721759 & 0.19210314 & 0.27773566 & 0.54870200 \\
 DDVTD & 0.153636 & 0.14637920 & 0.02640016 & 0.04233010 & 0.19171263 & 0.27376859 & 0.55795902 \\
 DDS   & 0.151186 & 0.14218154 & 0.03643847 & 0.08348558 & 0.13985555 & 0.23568086 & 0.34219700 \\
 DDST  & 0.153923 & 0.14673361 & $-3.786315 \cdot 10^{-5}$ & $1.611143 \cdot 10^{-5}$ & 0.13972293 & 0.20737662 & 0.56369799 \\
 DDSTD & 0.153999 & 0.14683193 & $-7.009164 \cdot 10^{-8}$ & 0.00000000 & 0.14036291 & 0.20810260 & 0.58325702\\
\noalign{\smallskip}\hline
\end{tabular}
\end{center}
\end{table*}

\begin{table*}
  \caption{Uncertainties of the observables used in the fitting of the
    parametrisations.}
\label{tab:03}
\begin{center}
\begin{tabular}{cccccc}
\hline\noalign{\smallskip}
parametrisation & binding & charge & diffraction & surface & spin-orbit \\
& energy & radius & radius & thickness & splitting \\
& uncertainty & uncertainty & uncertainty & uncertainty & uncertainty \\
& $\Delta O_{1}$ & $\Delta O_{2}$ & $\Delta O_{3}$ & $\Delta O_{4}$ & $\Delta O_{5}$ \\
& [MeV] & [fm]   & [fm]  & [fm] & [MeV] \\
\noalign{\smallskip}\hline\noalign{\smallskip}
 DDV   & 1.495500 & 0.011294 & 0.036662 & 0.023629 & 0.535689 \\
 DDVT  & 0.946271 & 0.013411 & 0.018134 & 0.026302 & 0.508166 \\
 DDVTD & 0.951520 & 0.013787 & 0.018455 & 0.027015 & 0.516422 \\
 DDS   & 1.612989 & 0.011381 & 0.045384 & 0.025996 & 0.529695 \\
 DDST  & 0.909146 & 0.013745 & 0.018559 & 0.026518 & 0.511721 \\
 DDSTD & 0.907443 & 0.014215 & 0.018800 & 0.027411 & 0.519149 \\
\noalign{\smallskip}\hline
\end{tabular}
\end{center}
\end{table*}

\clearpage

\begin{table*}
  \caption{Nuclear matter parameters of the models at saturation related to isoscalar properties with uncertainties in brackets.}
\label{tab:04}
\begin{center}
\begin{tabular}{ccccccc}
\hline\noalign{\smallskip}
parametrisation & mass of & Dirac & saturation & binding & incom- & skewness  \\
& $\sigma$ meson & effective & density & energy & pressibility & \\
& & mass & & per nucleon & &  \\
& $m_{\sigma}$ & $m^{\ast}/m_{\rm nuc}$ & $n_{\rm sat}$ & $a_{V}$ & $K$ & $Q$ \\
& [MeV] & & [fm$^{-3}$] & [MeV] & [MeV] & [MeV] \\
\noalign{\smallskip}\hline\noalign{\smallskip}
DDV   & 537.6(2.7) & 0.5863(0.0128) & 0.1511(0.0012)
& 16.28(0.06) & 240.0 & -610.6(339.5)   \\
DDVT  & 502.6(13.8) & 0.6668(0.0265) & 0.1536(0.0014)
& 16.27(0.04) & 240.0 & -743.3(391.2)   \\
DDVTD & 502.6(13.6) & 0.6671(0.0265) & 0.1536(0.0014)
& 16.27(0.06) & 240.0 & -763.4(403.8)   \\
DDS   & 539.3(2.9) & 0.5840(0.0156) & 0.1512(0.0014)
& 16.28(0.06) & 240.0 & -389.3(348.4)   \\
DDST  & 506.5(10.3) & 0.6716(0.0238) & 0.1539(0.0013)
& 16.28(0.04) & 240.0 & -735.7(257.2)   \\
DDSTD & 505.8(10.6) & 0.6730(0.0246) & 0.1540(0.0013)
& 16.28(0.04) & 240.0 & -737.6(221.3)   \\
\noalign{\smallskip}\hline
\end{tabular}
\end{center}
\end{table*}

\begin{table*}
  \caption{Nuclear matter parameters of the models at saturation
    related to isovector properties with uncertainties in brackets.}
\label{tab:05}
\begin{center}
\begin{tabular}{ccccc}
\hline\noalign{\smallskip}
parametrisation  & symmetry & symmetry & incom- & volume part \\
& energy & energy & pressibility & of isospin \\
& & slope & of symmetry & incom- \\
& & parameter & energy & pressibility \\
& $J$ & $L$ & $K_{\rm sym}$ & $K_{\tau,v}$\\
& [MeV] & [MeV] & [MeV] & [MeV]\\
\noalign{\smallskip}\hline\noalign{\smallskip}
 DDV   & 33.61(2.17) & 69.75(21.13) &  -97.53(28.62) & -338.6(93.3) \\
 DDVT  & 31.58(1.30) & 42.40(12.55) & -118.91(30.17) & -241.9(68.9) \\
 DDVTD & 31.57(1.34) & 42.58(13.49) & -116.58(32.57) & -236.6(71.1) \\
 DDS   & 33.98(2.39) & 74.58(22.29) &  -48.70(167.64) & -375.2(179.5) \\
 DDST  & 31.58(1.20) & 44.13(10.73) &  -85.33(436.54) & -214.8(455.0) \\
 DDSTD & 31.52(1.26) & 43.35(12.10) &  -82.97(906.46) & -209.8(940.2) \\
\noalign{\smallskip}\hline
\end{tabular}
\end{center}
\end{table*}

\clearpage

\clearpage

\clearpage

\end{document}